\documentclass[%
 reprint,
superscriptaddress,
 amsmath,amssymb,
 longbibliography,
aps,
pra,
]{revtex4-2}

\usepackage{graphicx}% Include figure files
\usepackage{bm}% bold math
\usepackage{xcolor}
\usepackage{hyperref}% add hypertext capabilities
\usepackage{orcidlink}
\usepackage{color}
\definecolor{BLUE}{rgb}{0.0,0.0,1.0}

\usepackage{soul}
\soulregister\cite7
\soulregister\ref7

\newcommand{\ket}[1]{| #1 \rangle}

\begin{document}

\title{Slowing YbF molecules using radiation pressure}

\author{M. Athanasakis-Kaklamanakis\orcidlink{0000-0003-0336-5980}}
\thanks{These authors contributed equally to this work.}

\author{G.~Peng\orcidlink{0009-0009-9583-3489
}}
\thanks{These authors contributed equally to this work.}

\author{S.~Li}
\author{H.~Septien-Gonzalez}
\author{C.~Debavelaere\orcidlink{0009-0003-8411-1575}}
\author{A.~D.~White\orcidlink{0009-0009-5038-8430}}
\author{S.~Popa}

\author{J.~Lim\orcidlink{0000-0002-1803-4642}}
\author{B.~E.~Sauer\orcidlink{0000-0002-3286-4853}}
\author{M.~R.~Tarbutt\orcidlink{0000-0003-2713-9531}}
\affiliation{Blackett Laboratory, Centre for Cold Matter, Imperial College London, SW7 2AZ London, United Kingdom}

\date{\today}

\begin{abstract}
We report radiation pressure slowing of YbF molecules to low velocity. In YbF, laser slowing is hindered by leaks out of the optical cycle attributed to low-lying metastable electronic states arising from inner-shell excitation. We bring this population back into the optical cycle once it has decayed to the electronic ground state using microwaves to couple the relevant rotational levels. We measure the scattering rate and closure of the optical cycle as repumps are added, and study the destabilzation of dark states by a magnetic field and by polarization modulation, finding that both are helpful for maximizing the scattering rate. Starting from a beam with a mean speed of 80~m/s, and using frequency broadened slowing light, we reduce the mean speed of the beam and produce a substantial flux in the low velocity tail of the distribution. Slowing increases the fraction of molecules below 40~m/s from 0.4(1)\% to 7.0(2)\%, and the fraction below 30~m/s from zero to 3.2(1)\%. The establishment of a nearly-closed optical cycle and the production of molecules at low velocity are important steps towards trapping YbF molecules for future measurements of the electron's electric dipole moment.

\end{abstract}


 \maketitle

\section{Introduction}
Ultracold polar molecules are versatile systems for quantum simulation of many-body phenomena~\cite{Cornish2024,Carr2009}, quantum computing~\cite{Holland2023, Bao2023}, ultracold chemistry experiments~\cite{Karman2024}, and tests of fundamental physics~\cite{Safronova2018}. The most precise measurements of the electron's electric dipole moment (eEDM, $d_e$) have all used heavy polar molecules~\cite{Hudson2011, Andreev2018, Roussy2023}, exploiting their large effective electric fields and high polarizability~\cite{Safronova2018,Cairncross2019,Arrowsmith-Kron2024}. The current limit, $|d_{e}| < 4.1\times10^{-30}~e$~cm~\cite{Roussy2023}, is sensitive to new physics at energies exceeding 10~TeV, probing hypothesized CP-violating interactions~\cite{Chupp2019} that are thought to be necessary to resolve the baryon asymmetry problem~\cite{Sakharov1991,Dine2003}. 

To improve on this limit, several groups aim to use ultracold polar molecules trapped in optical lattices~\cite{Fitch2020b,Anderegg2023,Bause2024arxiv, Zeng2024}, where high densities and long spin coherence times are possible. This requires species with high sensitivity to $d_e$ that can be produced at ultracold temperatures. An excellent candidate is ytterbium monofluoride (YbF). It is highly sensitive to the eEDM, parameterized by an effective electric field of $|E_{\rm{eff}}|\approx 26$~GV\,cm$^{-1}$~\cite{Kozlov1997} when fully polarized, it has been used to measure $d_e$ in the past~\cite{Hudson2002,Hudson2011}, and it is amenable to direct laser cooling~\cite{Tarbutt2013, Fitch2020b}. While the even isotopes are sensitive to eEDM, the odd isotopes can be used to search for CP violation in the ytterbium nucleus via the nuclear magnetic quadrupole moment~\cite{Ho2023} or the Schiff moment~\cite{Zheng2022}. Laser cooling has already been applied to a beam of YbF, reducing its temperature to about 100~$\mu$K~\cite{Lim2018, Alauze2021} in the two transverse directions. In those experiments, the forward velocity of the beam was about 200~m/s. To make progress, the beam needs to be decelerated to about 10~m/s so that it can be captured in a magneto-optical trap (MOT) and then transferred to an optical lattice. 

Radiation pressure slowing has been applied to several molecular species previously~\cite{Barry2012,Zhelyazkova2014,Truppe2017, Hemmerling2016}, but applying the technique to YbF is especially challenging. To our knowledge, the $A ^{2}\Pi_{1/2} \leftrightarrow X ^{2}\Sigma^{+}$ transition at 552~nm is the only viable cycling transition. Here, the single-photon recoil velocity is only 3.7~mm/s, whereas a typical cryogenic buffer gas beam has a mean speed of about 150~m/s. To bring this beam to rest, $4\times 10^4$ photons must be scattered and several vibrational repump lasers are needed to keep the molecules in the cooling cycle. After accounting for repumping, dark state destabilization, and finite laser power, photon scattering rates are limited to about $10^6$ photons/s. In these circumstances, laser slowing may take place over tens of milliseconds and several metres, with the consequent loss of flux from the beam divergence. These difficulties are compounded by the complex electronic structure of YbF, especially the presence of low-lying metastable electronic states~\cite{Zhang2022, Popa2024}, formed by excitation of inner-shell 4f electrons, known as the `4f hole' states.

Here, we demonstrate radiation pressure slowing of YbF to low velocity. As well as vibrational repuming, we use microwaves to repump population leaking to other rotational states of $X ^{2}\Sigma^{+}$ via the 4f hole states. We analyze potential bottlenecks in the photon cycling scheme and estimate the scattering rates as each repump is added. By broadening the spectrum of the slowing light to address the full range of velocities, and starting from a cryogenic buffer gas beam with mean speed around 80~m/s~\cite{White2024}, we show that molecules can be slowed to velocities close to the capture velocity of a magneto-optical trap.

\section{Methods}

\begin{figure*}[ht]%
\centering
\includegraphics[width=0.99\textwidth]{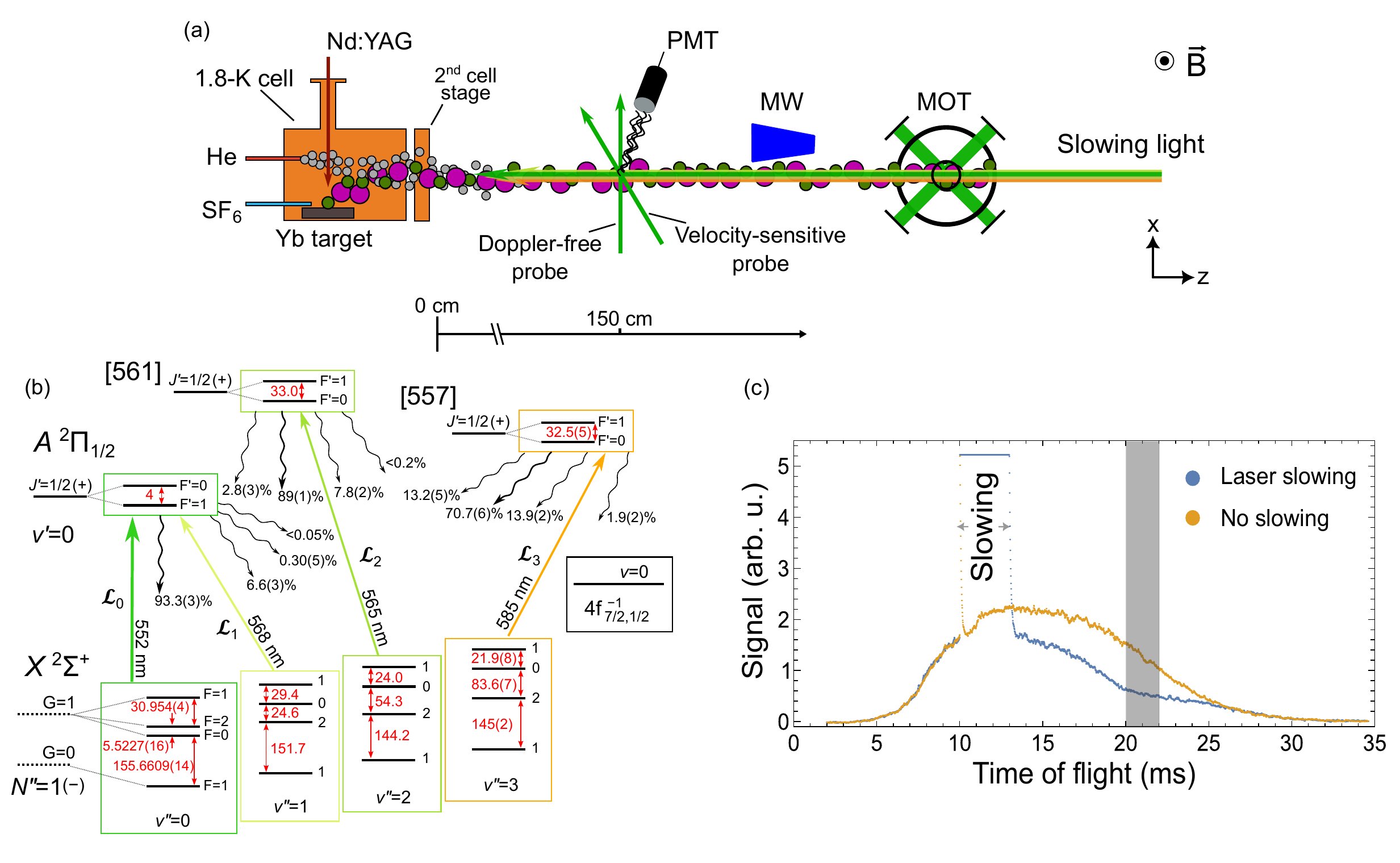}
\caption{Overview of laser slowing method. (a) Experimental setup. A beam of YbF is produced by a cryogenic buffer gas source. To slow the molecules, radiation pressure is applied by a counter-propagating laser beam and, optionally, microwave (MW) remixing. The microwaves are delivered to the molecules by an in-vacuum horn. The velocity distribution is determined from the Doppler shift between laser-induced fluorescence spectra recorded at two different angles. A uniform magnetic field is applied along $y$. (b) Energy levels of YbF used in this work showing the transitions driven by the slowing light and the branching ratios from the excited states. The hyperfine structure of each of the states is also shown.(c) Typical time-of-flight profile with and without laser slowing. The slowing laser is turned on between 10~ms and 12.7~ms. The detector is blinded by the slowing light during this period, but the rest of the profile is visible. When measuring the fraction of population remaining in $\ket{X;0;1}$, we typically use the shaded region of the profile.}\label{fig:setup}
\end{figure*}

To label rovibronic states succinctly, we use the notation $\ket{\eta;v;N}$, where $\eta$ identifies the electronic state ($X$ for $X^2\Sigma^+$, $A$ for $A^2\Pi_{1/2}$), $v$ is the vibrational quantum number, and $N$ is the rotational quantum number (where appropriate, $J$ is used). To identify specific hyperfine states, we use $\ket{\eta;v;N;F}$. When referring to a vibronic state without specifying a rotational state, the notation $\ket{\eta;v}$ is used. Two 4f hole states are also relevant in the present work and we label them $4f^{-1}_{7/2,1/2}$ and $4f^{-1}_{5/2,1/2}$,  following the notation used previously~\cite{Zhang2022, Popa2024}.

Figure~\ref{fig:setup}(a) shows the experimental setup used in this study. YbF molecules are produced by ablating a Yb target in the presence of He buffer gas mixed with SF$_6$ in a 1.8-K cryogenic cell. A two-stage cell design is used~\cite{White2024} and the He flow rate and ablation pulse energy are optimized to produce a beam of YbF with a peak velocity around 80~m/s and a velocity spread of about 50~m/s.

The molecules are detected by imaging laser-induced fluorescence (LIF) onto a photomultiplier tube labelled PMT in the figure. A laser beam drives the transition $\ket{A;0;1/2} \leftarrow \ket{X;0;1}$ at 552~nm, at an angle of either 90$^\circ$ (Doppler-free) or 60$^\circ$ (Doppler-sensitive) with respect to the molecular beam. The probe laser has a power of 4~mW and a $1/e^2$ diameter of 2.5~mm. Frequency sidebands at $-192$ and $-159$~MHz are applied using two acousto-optic modulators (AOM)  so that all hyperfine components of the transition are driven simultaneously. This produces a dominant, well-isolated, peak in the LIF spectrum, with a full width at half maximum (FWHM) of 32(1)~MHz in the Doppler-free spectrum. The Doppler-sensitive spectrum is converted to a velocity distribution, using the peak of the Doppler-free spectrum to determine zero velocity.

We slow down the molecules using the radiation pressure of a counter-propagating laser beam. Figure \ref{fig:setup}(b) shows the optical part of the cooling cycle. The main cooling transition is the rotationally-closed P(1) transition, $\ket{A;0;1/2} \leftarrow \ket{X;0;1}$, which is driven by laser $\mathcal{L}_{0}$ at 552~nm. This laser system consists of an external-cavity diode laser generating 30~mW at 1104~nm, which seeds a fiber amplifier whose output is then frequency doubled. Examining the branching ratios from $\ket{A,0}$ to $\ket{X,v''}$ shown in the figure, we see that it is necessary to repump population that leaks to $v''=1,2,3$ in order to scatter enough photons for effective slowing. This is achieved using the lasers labelled $\mathcal{L}_{v''}$. $\mathcal{L}_{1}$ drives $\ket{A;0;1/2}\leftarrow\ket{X;1;1}$. The state $\ket{A;1}$ is strongly mixed with $\ket{4f^{-1}_{5/2,1/2};0}$, leading to a pair of states labelled~\footnote{These labels specify the energy above the ground state, in THz. The same states are sometimes also called [18.58] and [18.71]~\cite{Popa2024} where the label is the energy in thousands of cm$^{-1}$.} as [561] and [557]~\cite{Lim2017}. These mixed states are used for repumping $v''=2$ and $v''=3$. The slowing laser beam has a $1/e^2$ diameter of 6.5~mm along $x$ and 5.9~mm along $y$, and powers of 600~mW for $\mathcal{L}_{0}$, 320~mW for $\mathcal{L}_{1}$, 140~mW for $\mathcal{L}_{2}$, and 15~mW for $\mathcal{L}_{3}$.  To address hyperfine structure, which is different for each vibrational state and is shown in Fig.~\ref{fig:setup}(b), a custom set of radio-frequency sidebands is applied to each laser using acousto-optic modulators (AOMs) and electro-optic modulators (EOMs). Zeeman dark states are destabilized by applying a magnetic field along $y$ and/or by modulating the polarizations of the slowing lasers using EOMs. The modulation frequency is 4.4~MHz for ${\cal L}_0$, and 6.5~MHz for $\mathcal{L}_{1}$, $\mathcal{L}_{2}$ and $\mathcal{L}_{3}$. The modulation depth is about 80\% for all lasers and is monitored by splitting the light on a polarizing beam cube and measuring the powers in each output. For experiments without polarization modulation, the light is linearly polarized in the $xy$-plane, at an angle that maximizes the photon scattering rate.

Figure \ref{fig:setup}(c) shows typical time-of-flight profiles recorded at the PMT by the Doppler-insensitive probe, with and without the laser slowing applied. Here, the slowing light is turned on from $t=10$~ms to $t=12.7$~ms. The slowing light blinds the PMT during this period, but it rapidly recovers. A fraction of the molecules that have seen the slowing light are lost, because the optical cycle is not fully closed as we will see later. We measure the population remaining in $\ket{X;0;1}$ in a selected time window, typically 20--22~ms, which is the shaded region in the figure. The frequencies of all the slowing lasers are tuned to be resonant with molecules arriving in this time window, accounting for the Doppler shift.

\section{Optical cycling}\label{sec:cycling}
Effective laser slowing requires a high photon scattering rate, which demands high excitation rates and the elimination of dark states. We follow a systematic approach to measuring and optimizing the scattering rate. First, we measure the scattering rate due to ${\cal L}_0$ alone, which we call ${\cal R}_0$. We do this by turning on ${\cal L}_0$ for a time $t_{\rm on}$, then measuring the fraction of molecules remaining in $\ket{X;0;1}$ as a function of $t_{\rm on}$ using the velocity-insensitive probe. This fraction decays exponentially with a $1/e$ time of $\tau_0$, and the scattering rate is ${\cal R}_0=\frac{1}{(1-r)\tau_0}$, where $1-r$ is the branching ratio for leaving the optical cycle which has been measured with a relative uncertainty  of 5\%~\cite{Zhuang2011}. 

\begin{figure}[tb]
    \centering
    \includegraphics[width=0.9\linewidth]{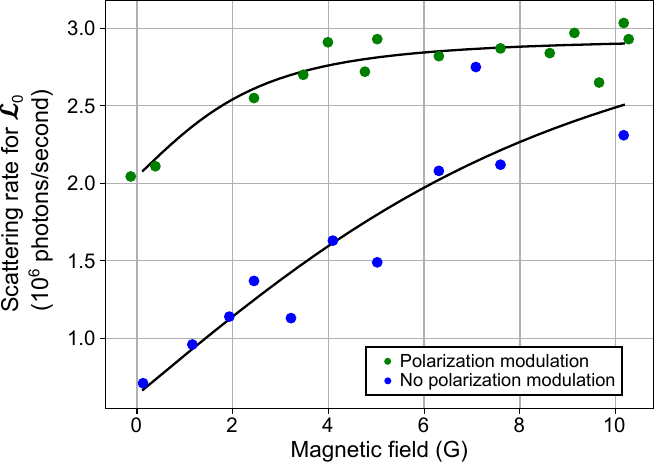}
    \caption{Photon scattering rate due to $\mathcal{L}_{0}$ alone, as a function of the applied magnetic field, $B$, with and without polarization modulation. The solid lines are fits to an empirical model: ${\cal R}_0(B) = {\cal R}_0(0) +{\cal R}_0'\frac{B}{B_{\rm s}} \left[ 1+ \left( \frac B{B_{\rm s}} \right) ^2 \right]^{-1/2}$.}
    \label{fig:polmod}
\end{figure}

 Figure \ref{fig:polmod} shows ${\cal R}_0$ as a function of applied magnetic field, $B$, with and without polarization modulation. The magnetic field destabilizes Zeeman dark states and does this optimally when $g \mu_{\rm B} B/\hbar \approx \Omega/4$~\cite{Berkeland2002}, where $\Omega$ is the Rabi frequency. If $B$ is too small the molecules spend too long in the dark state and if it is too large the scattering rate is reduced by Zeeman splitting. Unfortunately, in YbF the hyperfine components of $\ket{X;0;1}$ have very different magnetic $g$ factors, $g=0.071$ for the lower $F=1$ component, $g=0.43$ for the upper $F=1$ component, and $g=0.5$ for $F=2$. There can be Zeeman dark states in any of these hyperfine components so it is impossible to set the field optimally. Without polarization modulation, ${\cal R}_0$ increases for $B$ between 0 and 10~G. Note that we do not cancel the background magnetic field, which is why ${\cal R}_0$ does not drop to zero at $B=0$. For all values of $B$, the scattering rate increases when the polarization modulation is turned on. The polarization modulation should be effective in destabilizing all dark states without a magnetic field, but in practice ${\cal R}_0$ increases for $B$ between 0 and 4~G. It may be that the optimum polarization modulation rate ($\approx\Omega/2$) cannot be met for all hyperfine components because each has a different Rabi rate, so a combination of polarization modulation and magnetic field is optimal. In this case, we measure a maximum scattering rate of ${\cal R}_0 = 2.9(1)\times 10^6$ photons/s.
 
 In a multi-level system, where all transitions are driven on resonance at high enough intensity, with no dark states, the scattering rate approaches a maximum value 
\begin{equation}
{\cal R}^{\rm max}=\frac{n_e}{n_e+n_g}\Gamma,\label{eq:rmax}
\end{equation}
where $n_{g}$ and $n_{e}$ are the number of ground and excited states~\cite{Fitch2021b}, and $\Gamma = 2\pi \times 5.7 \times 10^6~\rm{s}^{-1}$ is the decay rate of the excited state. In our case, this is ${\cal R}^{\rm max}=\Gamma/4 = 9.0 \times 10^6$~photons/s. Thus, we obtain ${\cal R}_0 \approx {\cal R}^{\rm max}/3$ which is reasonable since finite laser intensities, detunings, and imperfect dark-state destabilization all lower scattering rates.

\begin{figure}[!ht]
    \centering
    \includegraphics[width=0.8\linewidth]{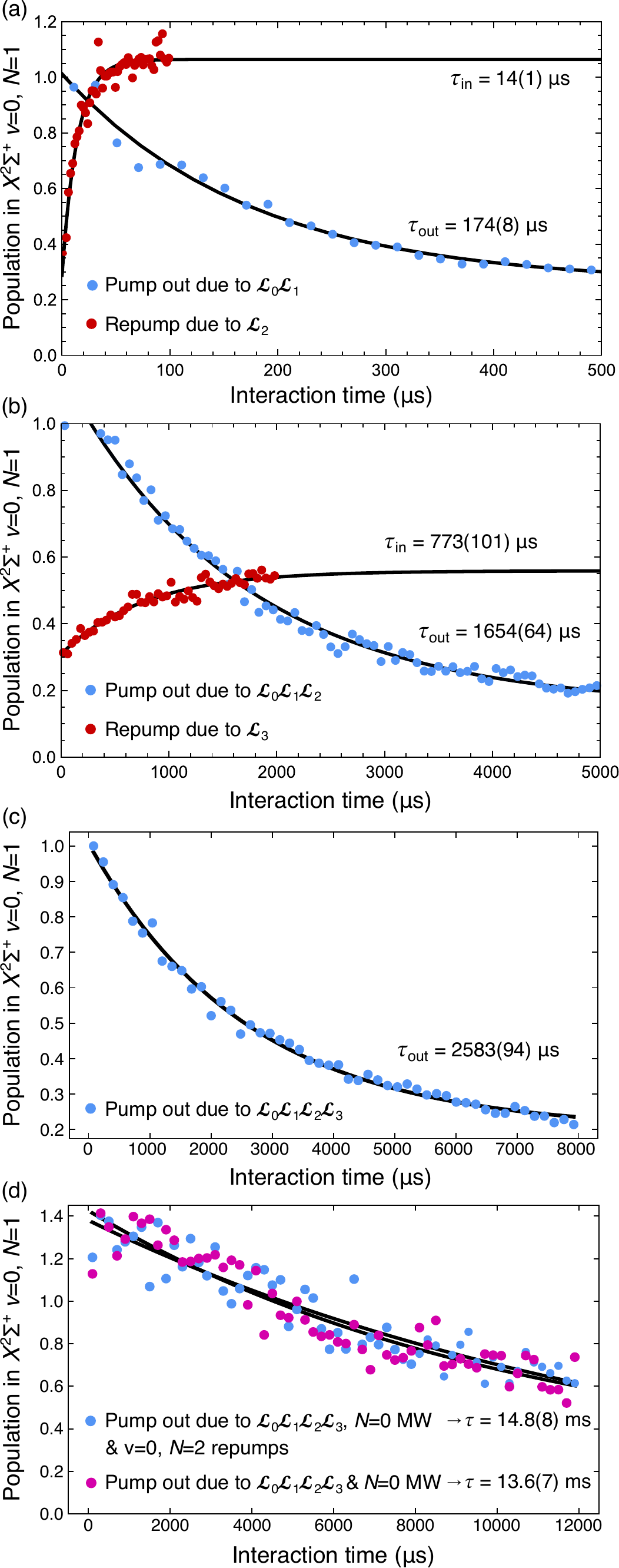}
    \caption{Timescales for pumping out of $\ket{X;0;1}$  (blue points), and pumping back in (red points) along with fits to an exponential decay (lines). In all cases, we use the time interval from 20-22~ms. (a) Decay of population in $\ket{X;0;1}$ due to pumping by ${\cal L}_0$ and ${\cal L}_1$ and recovery of population due to pumping by ${\cal L}_2$.  (b) Decay due to pumping by ${\cal L}_0$, ${\cal L}_1$ and ${\cal L}_2$, and recovery of population by ${\cal L}_3$. (c) Decay due to pumping by ${\cal L}_0$, ${\cal L}_1$, ${\cal L}_2$ and ${\cal L}_3$. (d) Decay due to pumping by ${\cal L}_0$, ${\cal L}_1$, ${\cal L}_2$ and ${\cal L}_3$, with microwave remixing to address population decaying to $\ket{X;v;0}$ for $v=0,1,2$ (purple points), as well as a laser repump to address molecules decaying to $\ket{X;0;2}$ (blue points). Here, we set the background to zero in the fits ($p_{\rm bg}=0$) in order to extract the decay times from the data.}
    \label{fig:repump_excitation_threeLevel}
\end{figure}

Next, we repeat this measurement with both ${\cal L}_0$ and ${\cal L}_1$ applied. The population decays more slowly, because the leak is reduced, and the same method is used to determine the new scattering rate which we call ${\cal R}_{0,1}$. As can be seen in Fig.~\ref{fig:setup}(b), ${\cal L}_1$ couples more ground states to the same excited state. Thus, following Eq.~(\ref{eq:rmax}), we expect ${\cal R}^{\rm max}$ to reduce to $\Gamma/7$, and the ratio of the scattering rates with and without ${\cal L}_1$ to be $4/7=0.571$. With polarization modulation applied to both lasers, we measure ${\cal R}_{0,1}/{\cal R}_0=0.59(1)$, consistent with this expectation. When dark states are destabilized by $B$ alone, we measure ${\cal R}_{0,1}/{\cal R}_0=0.48(1)$, showing again that this is a less effective method than polarization modulation.

Now we add ${\cal L}_2$ to the slowing light. This repump couples to a different excited state, so it need not reduce ${\cal R}_{\rm max}$, provided the excitation rate is high enough. We cannot measure the scattering rate, ${\cal R}_{0,1,2}$, using the same method as above because the uncertainty of the branching ratio out of the optical cycle is too large. Instead, we measure the timescale for pumping molecules out of $\ket{X;0;1}$ and $\ket{X;1;1}$ using $\mathcal{L}_{0}$ and $\mathcal{L}_{1}$, which we call $\tau_{\rm out}$, and the timescale for bringing population in $\ket{X;2;1}$ back to $\ket{X;0;1}$ using $\mathcal{L}_{1}$ and $\mathcal{L}_{2}$, which we call $\tau_{\rm in}$. These timescales are determined by measuring the population in $\ket{X;0;1}$ as a function of interaction time $t_{\rm{on}}$, and then fitting the results to a model of exponential decay towards equilibrium, $p(t)=p_{\rm bg}+a \exp(-t/\tau_{\rm in(out)})$. Figure \ref{fig:repump_excitation_threeLevel}(a) shows the results of these measurements, where we find $\tau_{\rm out}=174(8)$~$\mu$s and $\tau_{\rm in}=14(1)$~$\mu$s. In Appendix \ref{app:repump_model}, we describe a three-level model to characterize these optical pumping processes and show that the ratio of scattering rates is ${\cal R}_{0,1,2}/{\cal R}_{0,1}=1/(1+\tau_{\rm in}/\tau_{\rm out})$.
Our data yields the ratio $\tau_{\rm in}/\tau_{\rm out}=0.080(7)$, leading to ${\cal R}_{0,1,2}/{\cal R}_{0,1} = 0.926(6)$.

We use the same approach to determine the effect of ${\cal L}_3$, with the results shown in Fig.~\ref{fig:repump_excitation_threeLevel}(b). We find $\tau_{\rm in}/\tau_{\rm out}=0.47(6)$. In the case where the leak to $v=3$ is the dominant one, this analysis gives  a ratio of scattering rates of ${\cal R}_{0,1,2,3}/{\cal R}_{0,1,2} = 0.68(3)$. However, as we will soon see, there is another leak that is at least as large as the one to $v=3$. In this case, the above ratio is a lower limit because $\tau_{\rm out}$ is the timescale for pumping to any state outside the optical cycle, not just to $v=3$. Multiplying the factors together, we infer that the scattering rate with all slowing lasers applied lies in the range $1.58(6) \times 10^6 >{\cal R}_{0,1,2,3} > 1.08(6) \times 10^6$~photons/s.

Figure~\ref{fig:repump_excitation_threeLevel}(c) shows what happens when all four lasers ${\cal L}_0$, ${\cal L}_1$, ${\cal L}_2$, and ${\cal L}_3$ are used, all tuned to be resonant with molecules near 75~m/s. We see that the slowing light still depletes the population, now with a $1/e$ timescale of 2583(94)~$\mu$s, showing that there is still a leak out of the cooling cycle. Taking the decay time, together with the scattering rate determined above, we infer that the branching ratio out of the cooling cycle is between $2.5 \times 10^{-4}$ and $3.6\times 10^{-4}$. This is similar to the predicted branching ratio to $\ket{4f^{-1}_{7/2,1/2};0}$ ($4.2 \times 10^{-4}$) and to $\ket{4f^{-1}_{7/2,1/2};1}$ ($1.0 \times 10^{-4}$) \cite{Zhang2022}, so we suppose that this is the source of the loss. With such a leak, half the population is lost after slowing molecules by 7--10~m/s, so it is important to close this leak. We have attempted to do this in two different ways, as described next.

The $4f^{-1}_{7/2,1/2}$ electronic states lie between the $X$ and $A$ states and are metastable, having no direct electric dipole transition to the ground state. A quantum chemistry calculation estimates a lifetime of 8~ms~\cite{Zhang2022}. We first considered an optical approach to repump the population reaching these states. To search for a suitable transition, we used the slowing light to pump population into $\ket{4f^{-1}_{7/2,1/2};0}$, then drove the transition from this state back to $\ket{A;0}$,  using a CW laser at 1038~nm as the probe laser in the Doppler-free configuration at the PMT, and measured the 552~nm fluorescence due to the decay of $\ket{A;0}$. We scanned the frequency over several GHz, considerably wider than the uncertainty in the transition frequency determined from previous pulsed laser spectroscopy~\cite{Popa2024}. This search yielded a null result, and is described in more detail in Appendix~\ref{app:4f}. The absence of signal can be explained if the radiative lifetime of $4f^{-1}_{7/2,1/2}$ is considerably shorter than the predicted 8~ms, in which case the population decays to the ground state before we can detect it. 

\begin{figure}[tb]%
\centering
\includegraphics[width=0.95\linewidth]{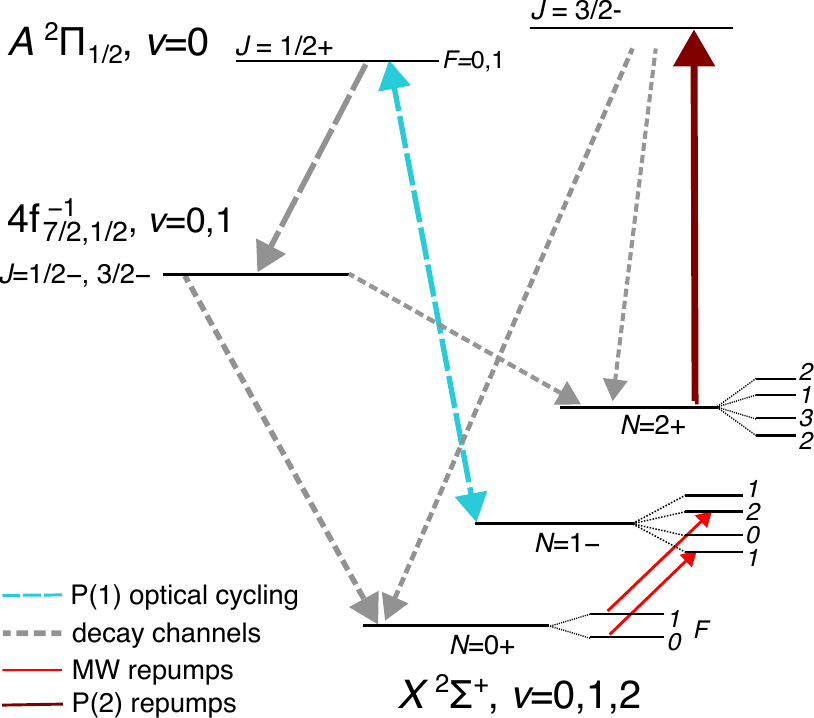}
\caption{Schematic of the relevant energy levels, laser repumps, and microwave remixing transitions used to bring molecules that decayed through the $4f^{-1}_{7/2,1/2}$ state back to the laser-cooling cycle.}\label{fig:microwave_remixing}
\end{figure}

\begin{figure}[tb]%
\centering
\includegraphics[width=0.95\linewidth]{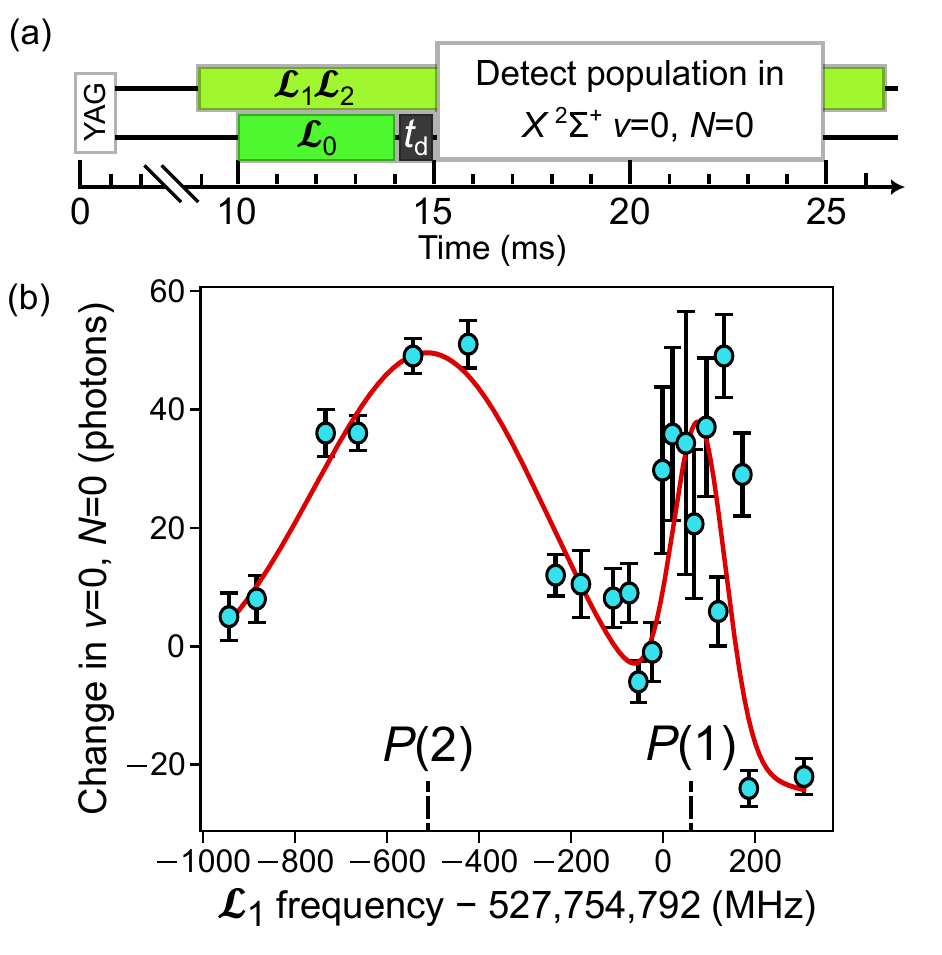}
\caption{Search for population leaking to $N=0$. (a) Timing sequence. (b) Change in population of $\ket{X;0;0}$ as a function of $\mathcal{L}_{1}$ frequency, while $\mathcal{L}_{0}$ and $\mathcal{L}_{2}$ drive the cycling transition. The red line is a fit to two Gaussian peaks plus a background with a linear gradient. The labels P(1) and P(2) indicate the expected frequencies for driving transitions from $N=1$ and $N=2$ respectively.}\label{fig:Q0}
\end{figure}

If the lifetime is short enough, it becomes possible to repump the population once it reaches the $X$ state. This was our next strategy. Figure \ref{fig:microwave_remixing} shows the potential decay pathways, based on the level structure  described in Ref.~\cite{Popa2024}. It is predicted that the leak primarily populates the negative parity components with $J=1/2$ and $J=3/2$ of $\ket{4f^{-1}_{7/2,1/2};0}$ and $\ket{4f^{-1}_{7/2,1/2};1}$. The leak to $\ket{4f^{-1}_{7/2,1/2};2}$ is estimated to be $10^{-6}$ so is negligible for our present purpose. When the population decays to $X$, it can only reach $N=0$ or $N=2$ due to parity and angular momentum selection rules, but several vibrational states may be populated. Table \ref{tab:FCF_4f_X} presents the relevant Franck-Condon factors which we have calculated using a harmonic oscillator model along with the bond lengths and vibrational constants given in Ref.~\cite{Zhang2022}. From this table, along with the prediction that the branching ratio into $\ket{4f^{-1}_{7/2,1/2};0}$ is 4.2 times that of $\ket{4f^{-1}_{7/2,1/2};1}$, we see that the decay chains will populate the $v=0,1,2$ states of $X$ with roughly equal probabilities, with smaller branches to higher $v$. 

To check that population is indeed reaching $N=0$, we applied the laser slowing light and monitored the population in $\ket{X;0;0}$ by driving the $\ket{A;0;1/2}\leftarrow \ket{X;0;0}$ transition (often called Q(0)). This measurement can be difficult to interpret because off-resonant excitation of nearby transitions by the laser slowing light also changes the rotational-state populations. This difficulty is exacerbated by the high intensity of the slowing light and the high spectral density of spectral lines in the vicinity of the cycling transition. To isolate the effect of interest from these background effects, we scan the frequency of ${\cal L}_1$ with the frequencies of the other lasers fixed. The source does not produce much population in $\ket{X;1}$, so off-resonant excitation by ${\cal L}_1$ has less impact. Figure~\ref{fig:Q0}(b) shows the change in the population of $\ket{X;0;0}$ as a function of the frequency of $\mathcal{L}_{1}$, using the experimental sequence shown in Figure \ref{fig:Q0}(a). We see a broad feature centred on the P(2) transition. This is due to molecules in $\ket{X;1;2}$ that are excited to $\ket{A;0;3/2}$ and then decay to $\ket{X;0;0}$. We see a narrower feature when $\mathcal{L}_{1}$ is resonant with the $P(1)$ transition, closing the optical cycle. We attribute this feature to the decay pathway into $N=0$ via the 4f hole states.

\begin{table}[tb]
\caption{Franck-Condon factors for transitions between $4f^{-1}_{7/2,1/2} (v')$ and $X^2\Sigma^+ (v'')$.} \label{tab:FCF_4f_X}
\begin{tabular}{p{1.5cm} p{1.5cm} p{1.5cm} p{1.5cm}}
\hline\hline
  & $v'=0$    & $v'=1$    & $v'=2$    \\
\hline
$v''=0$    & 0.2649 & 0.3131 & 0.2199\\
$v''=1$    & 0.3877 & 0.0273 & 0.0482\\
$v''=2$    & 0.2439 & 0.1146 & 0.1501\\
$v''=3$    & 0.0846 & 0.2855 & 0.0005\\
$v''=4$    & 0.0170 & 0.1906 & 0.1811\\
\hline\hline
\end{tabular}
\end{table}

One way to address the leaks to $N=0,2$ uses microwaves to drive $\ket{X;v;0}\leftrightarrow\ket{X;v;1}$ and $\ket{X;v;2}\leftrightarrow\ket{X;v;1}$. Unfortunately, according to Eq.~(\ref{eq:rmax}), this reduces the maximum possible scattering rate, ${\cal R}^{\rm max}$ from $\Gamma/7$ to $\Gamma/19$. A more efficient method uses microwaves to drive $\ket{X;v;0}\leftrightarrow\ket{X;v;1}$ and lasers to drive $\ket{A;0;3/2}\leftarrow \ket{X;0;2}$,  $\ket{A;0;3/2}\leftarrow \ket{X;1;2}$ etc.~\cite{Collopy2018}. This only reduces ${\cal R}^{\rm max}$ to $\Gamma/9$, so we prefer this method. 

We use microwaves to couple $\ket{X;v;1;1}\leftrightarrow\ket{X;v;0;0}$ and $\ket{X;v;1;2}\leftrightarrow\ket{X;v;0;1}$ for $v=0,1,2$, as illustrated in Fig.~\ref{fig:microwave_remixing}. The required set of frequencies are generated by mixing a local oscillator at 14362~MHz with an rf signal at 89 MHz, and applying frequency modulation at 800~kHz with a modulation index of 30 to the local oscillator. The microwaves are delivered to the molecules using an in-vacuum horn. The total power at the horn is 55~mW. The red points in Fig.~\ref{fig:repump_excitation_threeLevel}(d) show how the population decays when we apply all the slowing light and these microwaves. The microwaves increase the initial population because the $N=0$ populations are now part of the optical cycle and contribute to the signal at the detector. The decay time increases to 13.6(7)~ms, indicating that the microwaves are bringing the population back into the optical cycle which is now much closer to being closed. Next, to recover $N=2$ population, we add a laser repump to address $\ket{A;0;3/2}\leftarrow \ket{X;0;2}$. The result is shown by the blue points in Fig.~\ref{fig:repump_excitation_threeLevel}(d). This extra repump does not appear to make any significant difference. The pump out time is 14.8(8)~ms which is consistent with the previous decay time within the uncertainty. We suggest that off-resonant excitation by ${\cal L}_0$ is already sufficient to address $N=2$ at a high enough rate, since the intensity is high and the transition is only detuned by 736~MHz, making the additional repump unnecessary. In light of these findings, we use the $N=0$ (microwave) repumps but not the $N=2$ (laser) repumps for the remainder of this paper.

\section{Radiation pressure slowing}\label{sec:slowing}

\begin{figure}[tb]%
\centering
\includegraphics[width=0.9\linewidth]{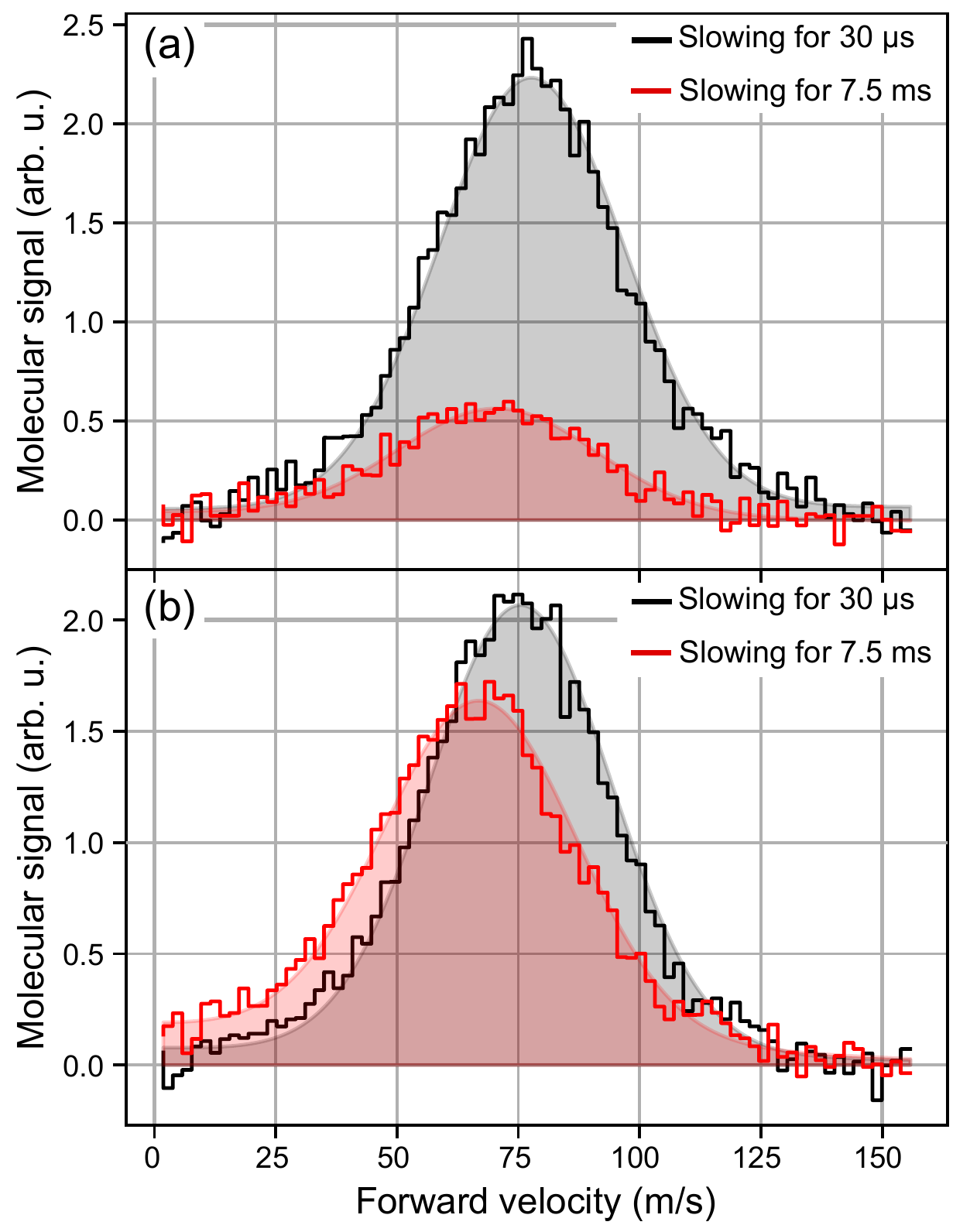}
\caption{Comparison of the velocity distribution of YbF molecules after slowing for 30~$\mu$s (black) and for 7.5~ms (red) when (a) only laser repumps are applied, and (b) microwaves are applied to bring $\ket{X;v;0}$ for $v=0,1,2$ into the laser-cooling cycle. The shaded curves are fits to Eq.~(\ref{eq:skewedGaussian}).}\label{fig:population_comparison}
\end{figure}

\begin{figure*}[ht]
    \centering
    \includegraphics[width=0.99\linewidth]{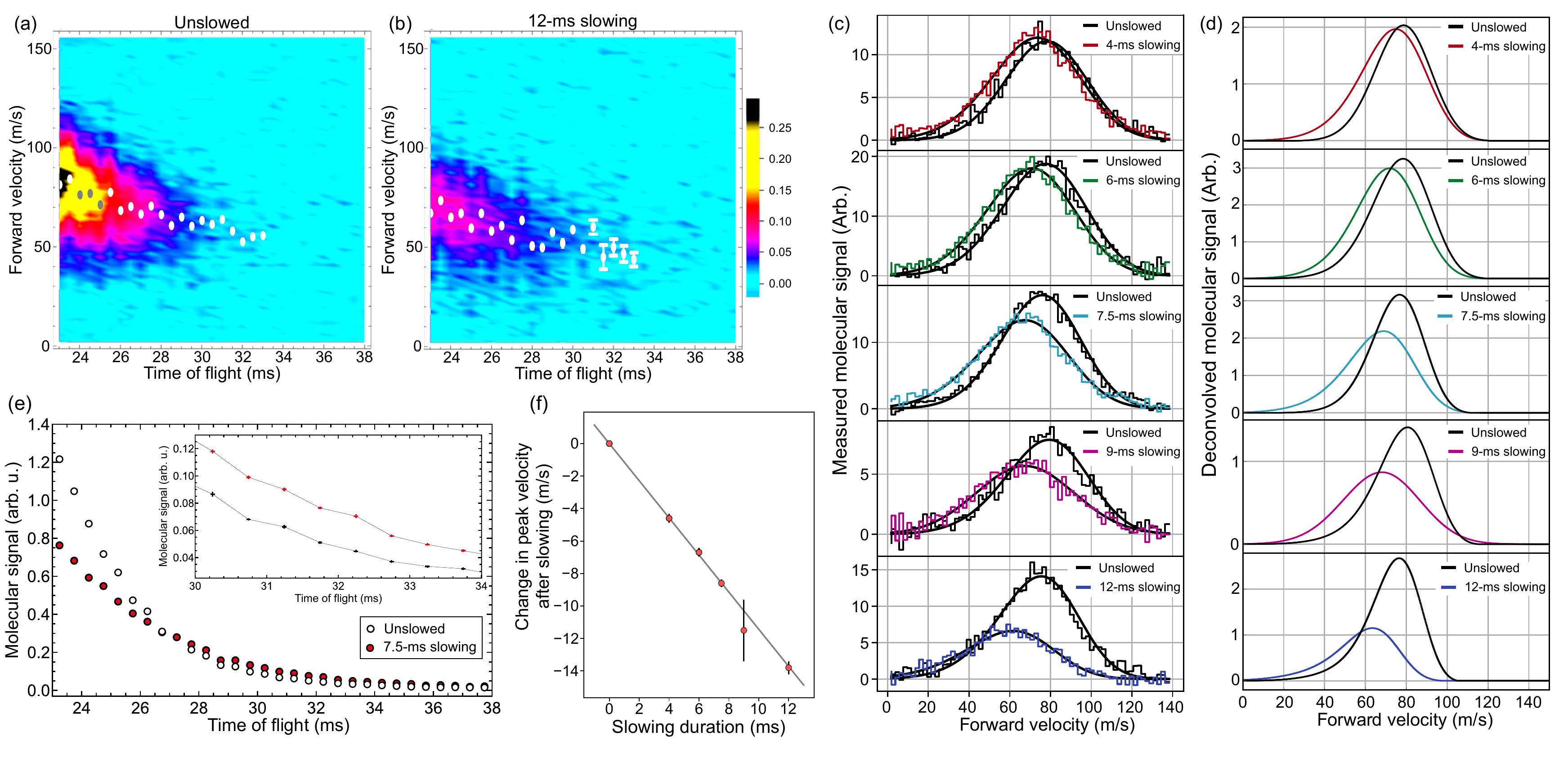}
    \caption{Radiation pressure slowing of YbF using frequency broadened light. (a,b) Intensity of LIF signal as a function of velocity and time of arrival at the detector. ${\cal L}_0$ is turned on at $t=10$~ms and held on for (a) $t_{\rm on}=30$~$\mu$s and (b) $t_{\rm on}=12$~ms. The other lasers and microwaves are on throughout. The intensity is in arbitrary units. The velocity distribution at each arrival time is determined from the distribution of Doppler shifts. The unslowed (or `control') distribution has $t_{\rm on}=30$~$\mu$s so as to capture any effects of population re-distribution by the slowing light.  The white dots mark the peak velocity in each 0.5-ms time-of-flight window. (c) Velocity distributions of molecules arriving between 23 and 38~ms, with and without laser slowing, for different slowing durations, $t_{\rm on}$. The lines are fits to Eq.~(\ref{eq:skewedGaussian}). (d) Velocity distributions obtained by deconvolving the instrumental profile from the fits in (c). (e) Time-of-flight distributions measured using Doppler-free LIF for $t_{\rm on}=30$~$\mu$s (open points) and $t_{\rm on}=7.5$~ms (filled points). The data shows the average of 400 shots and the error bars are smaller than the data-point markers. The inset shows a magnified view of the distribution beyond 30~ms. (f) Change in peak velocity as a function of $t_{\rm on}$. The line is a linear fit to the data and implies a constant acceleration of $a=-1143(8)$~m/s$^2$.}
    
    \label{fig:slowing}
\end{figure*}

To account for the changing Doppler shift as the molecules slow down, the frequency of the light can either be chirped or broadened~\cite{Barry2012,Zhelyazkova2014,Truppe2017, Hemmerling2016}. Here, we broaden the frequency spectra of each slowing laser to address all hyperfine components of the cycling transition for all velocities in a range of about 90~m/s. The methods used to do this are described in Appendix \ref{app:broadening}, and the resulting frequency spectra are shown in Fig.~\ref{fig:broadening}.

We measure the velocity distributions with and without slowing using the velocity-sensitive probe at the PMT. Figure~\ref{fig:population_comparison} shows the effect of introducing the microwave couplings illustrated in Fig.~\ref{fig:microwave_remixing}. We turn ${\cal L}_0$ on at $t=10$~ms and leave it on for $t_{\rm on}=30$~$\mu$s (black curves) and for $t_{\rm on}=7.5$~ms (red curves), first with microwaves off, then with microwaves on. The repump lasers (${\cal L}_{1},{\cal L}_{2}, {\cal L}_{3}$) are all on continuously. The distributions recorded with $t_{\rm on}=30$~$\mu$s act as control measurements, as this duration is long enough that rotational and hyperfine population re-distribution induced by the slowing light is also present in the control data, but short enough that any slowing is negligible. 

Without microwaves (see Fig.~\ref{fig:population_comparison}(a)), about 75\% of the molecules are lost due to the leak after slowing for 7.5~ms, and there is no increase in the number of molecules at any velocity. With microwaves (Fig.~\ref{fig:population_comparison}(b)), most of the molecules are recovered and there is a substantial increase in the low velocity wing of the distribution. We fit the distributions to a skewed Gaussian profile which has the form 
\begin{equation}
    f(v)\!=\!A\left(\!1\!+\! {\rm erf}\left[\frac{\gamma(v-v_0)}{\sqrt{2}\sigma }\right]\right)\exp\left[\!-\!\frac{(v-v_0)^2}{2\sigma^2} \right].
    \label{eq:skewedGaussian}
\end{equation}
The change in $v_0$ between the slowed and control distributions, $\Delta v_0$, is a convenient measure of the effectiveness of slowing. This is $\Delta v_0 = -7.2(6)$~m/s without microwaves, and $\Delta v_0 = -8.6(4)$~m/s with microwaves. Without microwaves, the mean speed of the remaining molecules is reduced, but there are not many of them, whereas with microwaves the mean speed is reduced by roughly the same amount, but far more remain.

Figure~\ref{fig:slowing} studies the slowing in more detail. As before, we measure a `slowed' distribution where ${\cal L}_0$ is turned on at $t=10$~ms for time $t_{\rm on}$, and a `control' distribution where it is only turned on for 30$~\mu$s. The other slowing lasers and all microwaves are on throughout. By scanning the probe laser and recording the time-of-flight profile at each frequency we obtain a two-dimensional map of the fluorescence intensity as a function of arrival time and velocity. Figure~\ref{fig:slowing}(a,b) show examples of these maps for the control data and slowed data with $t_{\rm on}=12$~ms. The main information in these maps is the central velocity as a function of arrival time, rather than the distribution of velocities in each time slice, since the latter is limited by the velocity resolution of the measurement. The white markers show these central velocities. Comparing the two maps, we see that in the slowed data the distribution has shifted both to lower velocity and towards later arrival times. 

Integrating over arrival time from 23 to 38~ms, we obtain the slowed velocity distribution, $S_{\rm slowed}(v)$, and the control velocity distribution, $S_{\rm control}(v)$. Figure \ref{fig:slowing}(c) shows these distributions for several values of $t_{\rm on}$, together with fits to Eq.~(\ref{eq:skewedGaussian}). 
We see that the peak of the slowed distribution shifts to lower velocity with increasing $t_{\rm on}$, with a substantial increase in the low velocity tail of the distribution. For $t_{\rm on}=4$~ms, the number of molecules detected increases when the slowing light is on because the deceleration shifts molecules from earlier arrival times into the 23-38~ms time window. For longer $t_{\rm on}$, deceleration of the beam is accompanied by a reduction in population. This may, in part, be due to unaddressed leaks to $N=2$, but it is also intrinsic: slower molecules drop further under gravity and diverge more rapidly, and this divergence is exacerbated by the transverse heating associated with photon scattering. As a result, the slower molecules are less likely to be detected meaning that the low velocity tail always underestimates the number of slow molecules produced. 

The distributions in Fig.~\ref{fig:slowing}(c) are obtained directly from the spectra measured using the velocity-sensitive probe. Each is a convolution of the actual velocity distribution in the beam with the instrumental profile arising due to other sources of spectral broadening. This instrumental profile can be determined using the orthogonal probe and is found to be a Gaussian distribution with a FWHM of 35~m/s. Figure \ref{fig:slowing}(d) shows the distributions obtained by deconvolving the fitted profiles in Fig.~\ref{fig:slowing}(c) from this instrumental profile, using the Richardson-Lucy algorithm. These profiles, which we label as $\tilde{ S}_{\rm{slowed}}$ and $\tilde{ S}_{\rm{control}}$, are our best estimates of the true velocity distributions for molecules arriving between 23 and 38~ms.

Figure \ref{fig:slowing}(e) shows the time-of-flight profiles beyond 23~ms, for  $t_{\rm on} = 7.5$~ms, measured using the velocity-insensitive probe and averaged over 400 shots. Slowing reduces the signal at times $t<27$~ms, and increases the signal at later times, as the slowed molecules take longer to reach the detector. Together, the velocity profiles and time-of-flight profiles provide clear evidence for radiation pressure slowing to low velocities. Although the low velocity tail is the most important part of the distribution for our purpose, we take the peaks of the velocity distributions as a convenient measure of the change in velocity. Figure \ref{fig:slowing}(f) shows this change as function of $t_{\rm on}$. The result fits well to a straight line, indicating a constant acceleration of $a=-1143(8)$~m/s$^2$. With this acceleration, molecules could be decelerated from 64~m/s to rest over the 1.8-m distance between source and MOT. The acceleration corresponds to a photon scattering rate of $3.06(2)\times10^5$ photons/s. This is an effective scattering rate averaged over the whole velocity distribution, and is lower than the rate determined in Sec.~\ref{sec:cycling} where we studied a narrow range of arrival times corresponding to molecules of the targeted velocity. 

\begin{figure}
    \centering
    \includegraphics[width=0.9\linewidth]{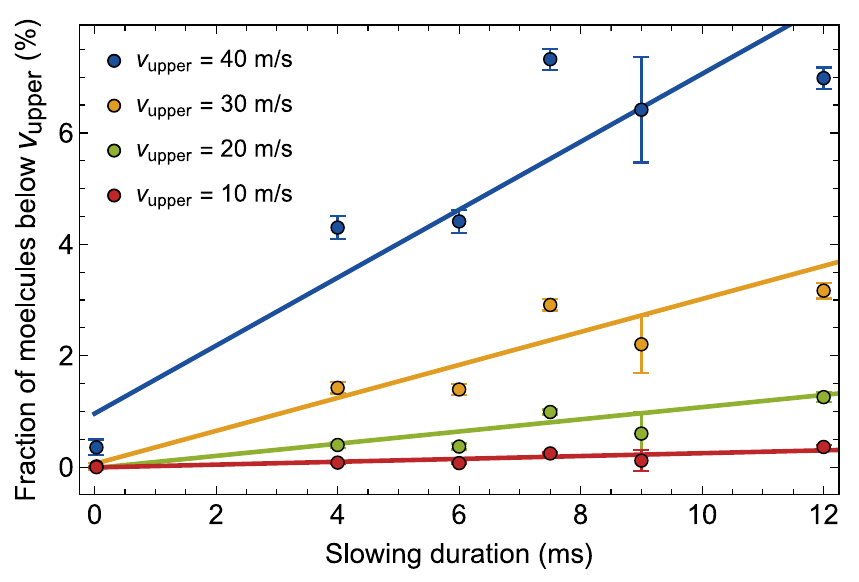}
    \caption{Fraction of YbF beam with a forward velocity below $v_{\rm{upper}}$ as a function of slowing duration ($t_{\rm on}$). The lines are linear fits to the data points, and are just guides to the eye.}
    \label{fig:fractionalMoleculesAsFuncOfSlowingDuration}
\end{figure}

From the Doppler-free LIF signal at the PMT, with no slowing applied, we estimate that there are $N_{\rm mol}=7(4)\times10^4$~molecules/shot passing through the detection volume in the time gate between 23 and 38~ms. The fraction in the slowed distribution with speeds below $v_{\rm upper}$ is defined as $r(v_{\rm{upper}}) = \int_0 ^{v_{\rm{upper}}} \tilde{S}_{\rm{slowed}}(v)dv/\int_0 ^{\infty} \tilde{S}_{\rm{control}}(v)dv$. Figure \ref{fig:fractionalMoleculesAsFuncOfSlowingDuration} shows this fraction versus $t_{\rm on}$, for several values of $v_{\rm{upper}}$. Here, the error bars are obtained by propagation of the uncertainties in the parameters of the fits to the data in Fig.~\ref{fig:slowing}(c). The lines are linear fits to guide the eye, though we do not expect $r(v_{\rm{upper}})$ to be a linear function of $t_{\rm on}$. Without slowing, the fraction below 40~m/s is 0.4(1)\% and the fraction below 30~m/s is consistent with zero. With slowing applied for 12~ms, the fraction below 40~m/s is 7.0(2)\%, the fraction below 30~m/s is 3.2(1)\%, and the fraction below 20~m/s is 1.3(1)\%. This corresponds to $9(5)\times10^{2}$~molecules/shot passing through the detection volume with velocity below 20~m/s.

Deceleration to these low velocities is an important step towards a MOT of YbF. Numerical simulations suggest that the MOT has a capture velocity of 10.5~m/s~\cite{Fitch2020b}, though this may be reduced by the lower scattering rate arising from the decay chain through the 4f hole states. We note that MOT beams typically have a diameter of at least 10~mm, whereas the $1/e^2$ diameter of the probe beam is only 2.5~mm. Thus the capture volume of the MOT is likely to be considerably larger than the detection volume used here, increasing our estimate for the number of slow molecules available for capture in the MOT.

\section{Summary and conclusions}

Laser cooling and trapping of heavy polar molecules such as YbF is challenging due to the low photon recoil velocity and the complex energy level structure. The optical cycle used successfully for other species in not fully closed for YbF because of intermediate electronic states arising from inner-shell excitation, known as the 4f hole states~\cite{Zhang2022,Popa2024}. The branching ratio to these states is approximately $3.6(2)\times 10^{-4}$, which is too large for effective slowing and trapping. An attempt to close this leak through optical repumping did not work, but we find that the lifetime of the relevant 4f hole state is short enough that the population can be recovered after it has decayed to the electronic ground state. To do this, we use microwaves to couple $N=0$ to $N=1$ within the three lowest vibrational states of $X^2\Sigma^+$. We have quantified the scattering rate as each repump is introduced, and have found that polarization modulation and an applied magnetic field are necessary to destabilize dark states most effectively. Starting from a beam with a mean speed near 80~m/s, we have demonstrated radiation-pressure slowing of YbF molecules to speeds below 20~m/s. We estimate that there are $9(5)\times10^2$~molecules/shot below this velocity after 12~ms of radiation-pressure slowing. This is a critical step towards capturing the molecules in a magneto-optical trap. We are currently introducing additional laser light to close any remaining leaks to $\ket{X;v>0;2}$, which may further improve the slowing. The slowing efficiency could also be improved by chirping the laser frequencies instead of broadening them~\cite{Truppe2017}, or by introducing a laser beam that co-propagates with the molecular beam to avoid the problem of over-slowing~\cite{Langin2023}, or by using a Zeeman slower~\cite{Petzold2018, Kaebert2021}. Once the molecules have been captured in a MOT, they can be cooled to sub-Doppler temperatures~\cite{Lim2018, Alauze2021} and transferred to an optical lattice. In the lattice, spin coherence times of several seconds are feasible, offering exceptional sensitivity to the electron's electric dipole moment and CP-violating electron-nucleon interactions~\cite{Fitch2020b}. By extending these methods to the fermionic isotopologues, $^{171}$YbF and $^{173}$YbF, it may be possible to study CP-violating effects within the nucleus such as the Schiff moment and the magnetic quadrupole moment~\cite{Ho2023}.

\begin{acknowledgements}
    We are grateful to Jon Dyne and David Pitman for their expert technical support, and to Jorge Mellado-Mu\~noz for earlier contributions supporting this work. This research has been funded in part by the Gordon and Betty Moore Foundation through Grants 8864 \& GBMF12327 (DOI 10.37807/GBMF12327), and by the Alfred P. Sloan Foundation under Grants G-2019-12505 \& G-2023-21035, and by UKRI under grants EP/X030180/1, ST/V00428X/1, ST/Y509978/1 and MR/Z505122/1.
\end{acknowledgements}

\appendix

\section{Influence of repumps on scattering rate}\label{app:repump_model}

\begin{figure}[tb]
\centering
\includegraphics[width=0.7\columnwidth]{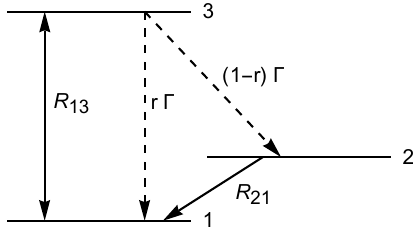}
\caption{Three-level model used to determine the influence of repumps on the photon scattering rate.}\label{fig:app:rate_model}
\end{figure}

To describe photon scattering, we consider the three-level rate model illustrated in Fig.~\ref{fig:app:rate_model}. Levels 1 and 3 form the main optical cycle and may represent multiple levels. Lasers couple these levels with excitation rate $R_{13}$. The population in 3 decays to 1 at rate $r \Gamma$ and to level 2 at rate $(1-r)\Gamma$, with $1-r \ll 1$. The population in 2 is pumped back to 1 at rate $R_{21}$. This re-pumping proceeds through a fourth state which is not part of the model and is a one-way process.

The populations $N_i(t)$ are described by the rate equations

\begin{equation}\label{eq:threelevel}
    \begin{split}
        \frac{dN_1(t)}{dt} & =R_{13}~ [N_3(t)-N_1(t)] +r~\Gamma N_3(t)+R_{21}~N_2(t),\\
       \frac{dN_2(t)}{dt} & =(1-r)~\Gamma ~ N_3(t)-R_{21}~N_2(t),\\
        \frac{dN_3(t)}{dt} & =R_{13} ~ [N_1(t)-N_3(t)] - \Gamma ~ N_3(t).
    \end{split}
\end{equation}

From the steady-state solutions, we obtain the photon scattering rate, ${\cal R}'=\Gamma N_3({\infty})$. This is
\begin{equation}
{\cal R}'=\frac{\Gamma}{(1-r)~\Gamma / R_{21}+2+\Gamma / R_{13}}.
\end{equation}
We can compare this to the scattering rate ${\cal R}$ obtained without the leak to level 2 (i.e. setting $r=1$). The ratio of these rates is
\begin{equation}
    \frac{{\cal R'}}{{\cal R}}=\frac{1}{1+\tau_{\rm in}/\tau_{\rm out}},
\end{equation}
where we have defined $\tau_{\rm in}=1/R_{21}$ and  $\tau_{\rm out}=\frac{2+\Gamma/R_{13}}{(1-r)\Gamma}$.

When $R_{13}=0$, $\tau_{\rm in}$ is the $1/e$ time scale for pumping population from 2 to 1. It can be determined by first pumping population into 2 (with the repump off), then turning the pump off and the repump on and measuring the population returned to 1 as a function of time. 

To understand the meaning of $\tau_{\rm out}$, we solve the rate equations with no repump ($R_{21}=0$). The result for the population remaining in the optical cycle is
\begin{equation}
1-N_2(t)=e^{-\frac{1}{2}pt}\left[\cosh\left(\frac{1}{2}q t\right)+\frac{p}{q}\sinh\left(\frac{1}{2}qt\right)\right],
\end{equation}
where
\begin{align}
    p &= 2R_{13}+\Gamma,\\
    q &= \sqrt{p^2-4xR_{13}\Gamma},
\end{align}
and we have introduced $x=1-r$. In the limit $x \ll 1$, we have $q \approx p + 2xR_{13}\Gamma/(2R_{13}+\Gamma)$, and the population remaining simplifies to
\begin{equation}
    1-N_2(t) \approx e^{-t/\tau_{\rm out}}.
\end{equation}
Thus, $\tau_{\rm out}$ can be determined by turning on the pump, with the repump left off ($R_{21}=0$), and measuring the population that remains as a function of time.

\section{Search for a repump transition from $4f^{-1}_{7/2,1/2}~(v=0)$} \label{app:4f}

\begin{figure*}[tb]%
\centering
\includegraphics[width=0.99\textwidth]{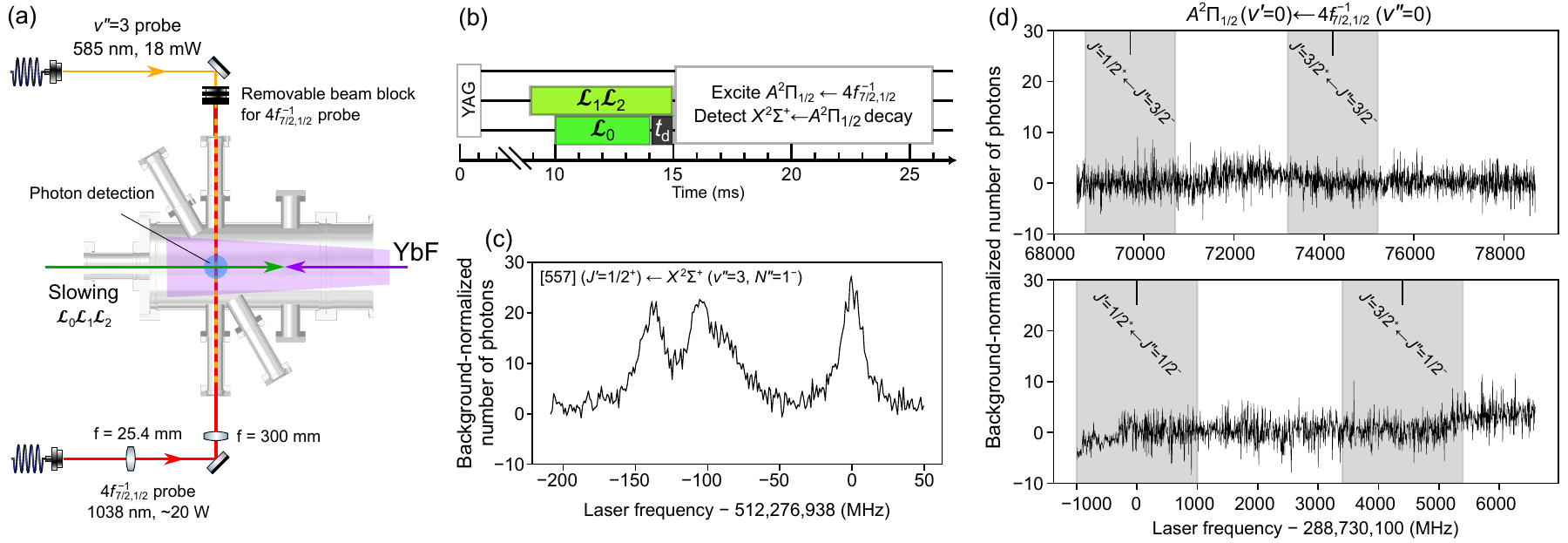}
\caption{Search for repump transitions for molecules that decay to the metastable $4f^{-1}_{7/2,1/2}$ state. (a)~Experimental apparatus. (b)~Timing sequence. (c)~Fluorescence obtained by probing the molecules that decayed to $\ket{X;3}$. (d)~Search for fluorescence from molecules repumped from $4f^{-1}_{7/2,1/2}$. }\label{fig:4f_search}
\end{figure*}

In the main text, we discussed the leak out of the optical cycle due to the 4f hole states. If the lifetime of these metastable states is too long, the only approach to recovering the population and maintaining a high scattering rate is to repump the population directly from the relevant 4f hole states. To this end, we searched for the repump transition $\ket{A;0} \leftarrow \ket{4f^{-1}_{7/2,1/2};0}$. The required frequency is known with an uncertainty of $270$~MHz from earlier pulsed laser spectroscopy of other transitions involving these 4f hole states~\cite{Popa2024}. 

Figure \ref{fig:4f_search}(a) shows the experimental setup used for this search. The slowing lasers $\mathcal{L}_{0}$, $\mathcal{L}_{1}$, and $\mathcal{L}_{2}$ are pulsed on for $t_{\rm on}=5$~ms in order to pump population into $\ket{4f^{-1}_{7/2,1/2};0}$. The molecules then pass through a probe laser that drives the $\ket{A;0} \leftarrow \ket{4f^{-1}_{7/2,1/2};0}$ transition at 1038~nm. This probe has a power of 20~W and a linewidth of 100~kHz. $\ket{A;0}$ decays promptly to the ground electronic state, and we detect the resulting fluorescence at 552~nm on a PMT. An interference filter is used to block background light at other wavelengths. However, ${\cal L}_0$ blinds the PMT, so we wait for a dead-time of $t_{\rm{d}}=1$~ms before recording the signal. To remove residual background, we turn the laser slowing light on and off in consecutive shots, and subtract one dataset from the other.

Quantum chemistry calculations~\cite{Zhang2022} suggest that the branching ratio for $\ket{A;0} \rightarrow \ket{4f^{-1}_{7/2,1/2};0}$ is four times larger than for $\ket{A;0}\rightarrow\ket{X;3}$. Thus, we can record the population reaching $\ket{X;3}$ in order to verify that the optical pumping is working as intended and to calibrate the expected signal size. To this end, we introduce a second orthogonal probe laser that drives $[557] \leftarrow \ket{X;3}$ at 585~nm, and detect the resulting fluorescence. Figure \ref{fig:4f_search}(c) shows an example of a spectrum recorded on this transition. Based on the predicted branching ratios, we could expect the signal from the 4f hole search to be about 4 times larger and thus easily detectable.

Figure \ref{fig:4f_search}(d) shows the results of this search. The optical pumping can prepare population in the negative parity component of $\ket{4f^{-1}_{7/2,1/2};0;1/2}$ or $\ket{4f^{-1}_{7/2,1/2};0;3/2}$, and these could be driven to the positive parity component of $\ket{A;0;1/2}$ or $\ket{A;0;3/2}$. We scanned the probe laser over all four spectral regions. The gray bands in Fig.~\ref{fig:4f_search}(d) denote a $\pm$1~GHz interval around each of the expected transition frequencies. We found no statistically significant evidence of any resonant fluorescence induced by the probe laser. We surmise that most of the molecules optically pumped into the 4f hole states decay before they reach the probe laser, implying a lifetime substantially shorter than the calculated value~\cite{Zhang2022} of 8~ms.
\\
\section{Frequency broadening of slowing lasers}\label{app:broadening}

\begin{figure*}[t]%
\centering
\includegraphics[width=0.85\textwidth]{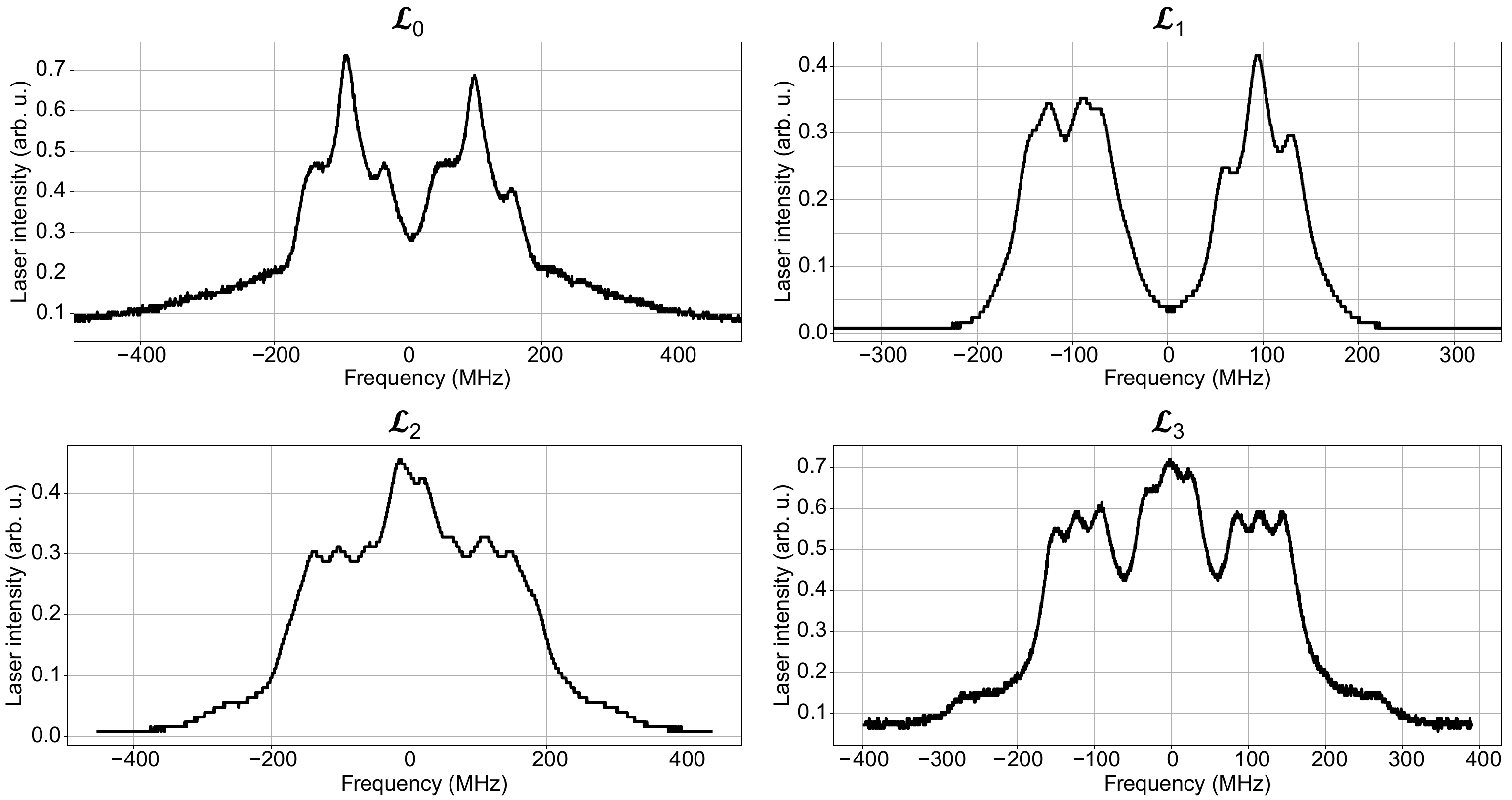}
\caption{Frequency spectra of the spectrally broadened lasers used for laser slowing of YbF, measured with a Fabry-Per\'ot cavity.}\label{fig:broadening}
\end{figure*}

We shape the spectrum of each slowing laser in order to address all hyperfine components of the cycling transition, and a continuous range of velocities spanning about 90~m/s. In the following, we describe the setup used for each laser. Figure \ref{fig:broadening} shows the resulting frequency spectra, measured using an optical cavity with a linewidth of about 10~MHz.

To broaden $\mathcal{L}_{0}$, we use a fiber-based EOM between the seed laser and the amplifier, and apply symmetric and roughly equal first-order sidebands at $\pm$13, 25, 35, 47, 61, and 75~MHz that persist after amplification and second-harmonic generation. After doubling, this broadened light is split in two, the frequency of one component is shifted by $-192$~MHz using an AOM in order to span the hyperfine components, the two parts are recombined, and the beam sent through the polarization modulation EOM. The resulting spectrum consists of two broad peaks spaced by 192~MHz, each about 160~MHz wide.

For $\mathcal{L}_{1}$, the frequency spectrum is shaped by splitting the laser output into two beams, passing one through an AOM at $-176$~MHz followed by an EOM driven at 35~MHz, passing the other through a 26 MHz free-space EOM, and then recombining the two beams. The 26 MHz EOM is driven with high rf power to generate the first, second and third-order sidebands, while the 35 MHz EOM generates only first-order sidebands.

For $\mathcal{L}_{2}$ the required spectrum is generated by passing the light through two free-space EOMs in succession, one at 111~MHz and the other at 33~MHz. For $\mathcal{L}_{3}$, we use free-space EOMs at 115~MHz and 31~MHz. All are driven to produce only first-order sidebands.

Before entering the beamline, $\mathcal{L}_{1}$, $\mathcal{L}_{2}$, and $\mathcal{L}_{3}$ pass through a polarization-modulation EOM at 6.5~MHz, which is driven with high rf power such that it produces roughly equal first-, second-, and third-harmonic frequency sidebands at $\pm$6.5, $\pm$13, and $\pm$19.5~MHz, respectively. These sidebands ensure that the broadened frequency spectra of the repumps are dense and uniform, with the spacing between successive frequency components being close to the natural linewidth of the transitions they address.

\bibliography{references}

\begin{thebibliography}{41}%
\makeatletter
\providecommand \@ifxundefined [1]{%
 \@ifx{#1\undefined}
}%
\providecommand \@ifnum [1]{%
 \ifnum #1\expandafter \@firstoftwo
 \else \expandafter \@secondoftwo
 \fi
}%
\providecommand \@ifx [1]{%
 \ifx #1\expandafter \@firstoftwo
 \else \expandafter \@secondoftwo
 \fi
}%
\providecommand \natexlab [1]{#1}%
\providecommand \enquote  [1]{``#1''}%
\providecommand \bibnamefont  [1]{#1}%
\providecommand \bibfnamefont [1]{#1}%
\providecommand \citenamefont [1]{#1}%
\providecommand \href@noop [0]{\@secondoftwo}%
\providecommand \href [0]{\begingroup \@sanitize@url \@href}%
\providecommand \@href[1]{\@@startlink{#1}\@@href}%
\providecommand \@@href[1]{\endgroup#1\@@endlink}%
\providecommand \@sanitize@url [0]{\catcode `\\12\catcode `\$12\catcode
  `\&12\catcode `\#12\catcode `\^12\catcode `\_12\catcode `\%12\relax}%
\providecommand \@@startlink[1]{}%
\providecommand \@@endlink[0]{}%
\providecommand \url  [0]{\begingroup\@sanitize@url \@url }%
\providecommand \@url [1]{\endgroup\@href {#1}{\urlprefix }}%
\providecommand \urlprefix  [0]{URL }%
\providecommand \Eprint [0]{\href }%
\providecommand \doibase [0]{https://doi.org/}%
\providecommand \selectlanguage [0]{\@gobble}%
\providecommand \bibinfo  [0]{\@secondoftwo}%
\providecommand \bibfield  [0]{\@secondoftwo}%
\providecommand \translation [1]{[#1]}%
\providecommand \BibitemOpen [0]{}%
\providecommand \bibitemStop [0]{}%
\providecommand \bibitemNoStop [0]{.\EOS\space}%
\providecommand \EOS [0]{\spacefactor3000\relax}%
\providecommand \BibitemShut  [1]{\csname bibitem#1\endcsname}%
\let\auto@bib@innerbib\@empty
%</preamble>
\bibitem [{\citenamefont {Cornish}\ \emph {et~al.}(2024)\citenamefont
  {Cornish}, \citenamefont {Tarbutt},\ and\ \citenamefont
  {Hazzard}}]{Cornish2024}%
  \BibitemOpen
  \bibfield  {author} {\bibinfo {author} {\bibfnamefont {S.~L.}\ \bibnamefont
  {Cornish}}, \bibinfo {author} {\bibfnamefont {M.~R.}\ \bibnamefont
  {Tarbutt}},\ and\ \bibinfo {author} {\bibfnamefont {K.~R.~A.}\ \bibnamefont
  {Hazzard}},\ }\bibfield  {title} {\bibinfo {title} {Quantum computation and
  quantum simulation with ultracold molecules},\ }\href
  {https://doi.org/10.1038/s41567-024-02453-9} {\bibfield  {journal} {\bibinfo
  {journal} {Nat. Phys.}\ }\textbf {\bibinfo {volume} {20}},\ \bibinfo {pages}
  {730} (\bibinfo {year} {2024})}\BibitemShut {NoStop}%
\bibitem [{\citenamefont {Carr}\ \emph {et~al.}(2009)\citenamefont {Carr},
  \citenamefont {DeMille}, \citenamefont {Krems},\ and\ \citenamefont
  {Ye}}]{Carr2009}%
  \BibitemOpen
  \bibfield  {author} {\bibinfo {author} {\bibfnamefont {L.~D.}\ \bibnamefont
  {Carr}}, \bibinfo {author} {\bibfnamefont {D.}~\bibnamefont {DeMille}},
  \bibinfo {author} {\bibfnamefont {R.~V.}\ \bibnamefont {Krems}},\ and\
  \bibinfo {author} {\bibfnamefont {J.}~\bibnamefont {Ye}},\ }\bibfield
  {title} {\bibinfo {title} {{Cold and ultracold molecules: science, technology
  and applications}},\ }\bibfield  {journal} {\bibinfo  {journal} {New J.
  Phys.}\ }\textbf {\bibinfo {volume} {11}},\ \href
  {https://doi.org/10.1088/1367-2630/11/5/055049}
  {10.1088/1367-2630/11/5/055049} (\bibinfo {year} {2009})\BibitemShut
  {NoStop}%
\bibitem [{\citenamefont {Holland}\ \emph {et~al.}(2023)\citenamefont
  {Holland}, \citenamefont {Lu},\ and\ \citenamefont {Cheuk}}]{Holland2023}%
  \BibitemOpen
  \bibfield  {author} {\bibinfo {author} {\bibfnamefont {C.~M.}\ \bibnamefont
  {Holland}}, \bibinfo {author} {\bibfnamefont {Y.}~\bibnamefont {Lu}},\ and\
  \bibinfo {author} {\bibfnamefont {L.~W.}\ \bibnamefont {Cheuk}},\ }\bibfield
  {title} {\bibinfo {title} {On-demand entanglement of molecules in a
  reconfigurable optical tweezer array},\ }\href
  {https://doi.org/10.1126/science.adf4272} {\bibfield  {journal} {\bibinfo
  {journal} {Science}\ }\textbf {\bibinfo {volume} {382}},\ \bibinfo {pages}
  {1143} (\bibinfo {year} {2023})}\BibitemShut {NoStop}%
\bibitem [{\citenamefont {Bao}\ \emph {et~al.}(2023)\citenamefont {Bao},
  \citenamefont {Yu}, \citenamefont {Anderegg}, \citenamefont {Chae},
  \citenamefont {Ketterle}, \citenamefont {Ni},\ and\ \citenamefont
  {Doyle}}]{Bao2023}%
  \BibitemOpen
  \bibfield  {author} {\bibinfo {author} {\bibfnamefont {Y.}~\bibnamefont
  {Bao}}, \bibinfo {author} {\bibfnamefont {S.~S.}\ \bibnamefont {Yu}},
  \bibinfo {author} {\bibfnamefont {L.}~\bibnamefont {Anderegg}}, \bibinfo
  {author} {\bibfnamefont {E.}~\bibnamefont {Chae}}, \bibinfo {author}
  {\bibfnamefont {W.}~\bibnamefont {Ketterle}}, \bibinfo {author}
  {\bibfnamefont {K.-K.}\ \bibnamefont {Ni}},\ and\ \bibinfo {author}
  {\bibfnamefont {J.~M.}\ \bibnamefont {Doyle}},\ }\bibfield  {title} {\bibinfo
  {title} {Dipolar spin-exchange and entanglement between molecules in an
  optical tweezer array},\ }\href {https://doi.org/10.1126/science.adf8999}
  {\bibfield  {journal} {\bibinfo  {journal} {Science}\ }\textbf {\bibinfo
  {volume} {382}},\ \bibinfo {pages} {1138} (\bibinfo {year}
  {2023})}\BibitemShut {NoStop}%
\bibitem [{\citenamefont {Karman}\ \emph {et~al.}(2024)\citenamefont {Karman},
  \citenamefont {Tomza},\ and\ \citenamefont {P\'erez-R\'ios}}]{Karman2024}%
  \BibitemOpen
  \bibfield  {author} {\bibinfo {author} {\bibfnamefont {T.}~\bibnamefont
  {Karman}}, \bibinfo {author} {\bibfnamefont {M.}~\bibnamefont {Tomza}},\ and\
  \bibinfo {author} {\bibfnamefont {J.}~\bibnamefont {P\'erez-R\'ios}},\
  }\bibfield  {title} {\bibinfo {title} {Ultracold chemistry as a testbed for
  few-body physics},\ }\href {https://doi.org/10.1038/s41567-024-02467-3}
  {\bibfield  {journal} {\bibinfo  {journal} {Nature Physics}\ }\textbf
  {\bibinfo {volume} {20}},\ \bibinfo {pages} {722} (\bibinfo {year}
  {2024})}\BibitemShut {NoStop}%
\bibitem [{\citenamefont {Safronova}\ \emph {et~al.}(2018)\citenamefont
  {Safronova}, \citenamefont {Budker}, \citenamefont {DeMille}, \citenamefont
  {Kimball}, \citenamefont {Derevianko},\ and\ \citenamefont
  {Clark}}]{Safronova2018}%
  \BibitemOpen
  \bibfield  {author} {\bibinfo {author} {\bibfnamefont {M.~S.}\ \bibnamefont
  {Safronova}}, \bibinfo {author} {\bibfnamefont {D.}~\bibnamefont {Budker}},
  \bibinfo {author} {\bibfnamefont {D.}~\bibnamefont {DeMille}}, \bibinfo
  {author} {\bibfnamefont {D.~F.~J.}\ \bibnamefont {Kimball}}, \bibinfo
  {author} {\bibfnamefont {A.}~\bibnamefont {Derevianko}},\ and\ \bibinfo
  {author} {\bibfnamefont {C.~W.}\ \bibnamefont {Clark}},\ }\bibfield  {title}
  {\bibinfo {title} {{Search for new physics with atoms and molecules}},\
  }\href {https://doi.org/10.1103/RevModPhys.90.025008} {\bibfield  {journal}
  {\bibinfo  {journal} {Rev. Mod. Phys.}\ }\textbf {\bibinfo {volume} {90}},\
  \bibinfo {pages} {025008} (\bibinfo {year} {2018})}\BibitemShut {NoStop}%
\bibitem [{\citenamefont {Hudson}\ \emph {et~al.}(2011)\citenamefont {Hudson},
  \citenamefont {Kara}, \citenamefont {Smallman}, \citenamefont {Sauer},
  \citenamefont {Tarbutt},\ and\ \citenamefont {Hinds}}]{Hudson2011}%
  \BibitemOpen
  \bibfield  {author} {\bibinfo {author} {\bibfnamefont {J.~J.}\ \bibnamefont
  {Hudson}}, \bibinfo {author} {\bibfnamefont {D.~M.}\ \bibnamefont {Kara}},
  \bibinfo {author} {\bibfnamefont {I.~J.}\ \bibnamefont {Smallman}}, \bibinfo
  {author} {\bibfnamefont {B.~E.}\ \bibnamefont {Sauer}}, \bibinfo {author}
  {\bibfnamefont {M.~R.}\ \bibnamefont {Tarbutt}},\ and\ \bibinfo {author}
  {\bibfnamefont {E.~A.}\ \bibnamefont {Hinds}},\ }\bibfield  {title} {\bibinfo
  {title} {Improved measurement of the shape of the electron},\ }\href
  {https://doi.org/10.1038/nature10104} {\bibfield  {journal} {\bibinfo
  {journal} {Nature}\ }\textbf {\bibinfo {volume} {473}},\ \bibinfo {pages}
  {493} (\bibinfo {year} {2011})}\BibitemShut {NoStop}%
\bibitem [{\citenamefont {Andreev}\ \emph {et~al.}(2018)\citenamefont
  {Andreev}, \citenamefont {Ang}, \citenamefont {DeMille}, \citenamefont
  {Doyle}, \citenamefont {Gabrielse}, \citenamefont {Haefner}, \citenamefont
  {Hutzler}, \citenamefont {Lasner}, \citenamefont {Meisenhelder},
  \citenamefont {O’Leary}, \citenamefont {Panda}, \citenamefont {West},
  \citenamefont {West},\ and\ \citenamefont {Wu}}]{Andreev2018}%
  \BibitemOpen
  \bibfield  {author} {\bibinfo {author} {\bibfnamefont {V.}~\bibnamefont
  {Andreev}}, \bibinfo {author} {\bibfnamefont {D.~G.}\ \bibnamefont {Ang}},
  \bibinfo {author} {\bibfnamefont {D.}~\bibnamefont {DeMille}}, \bibinfo
  {author} {\bibfnamefont {J.~M.}\ \bibnamefont {Doyle}}, \bibinfo {author}
  {\bibfnamefont {G.}~\bibnamefont {Gabrielse}}, \bibinfo {author}
  {\bibfnamefont {J.}~\bibnamefont {Haefner}}, \bibinfo {author} {\bibfnamefont
  {N.~R.}\ \bibnamefont {Hutzler}}, \bibinfo {author} {\bibfnamefont
  {Z.}~\bibnamefont {Lasner}}, \bibinfo {author} {\bibfnamefont
  {C.}~\bibnamefont {Meisenhelder}}, \bibinfo {author} {\bibfnamefont {B.~R.}\
  \bibnamefont {O’Leary}}, \bibinfo {author} {\bibfnamefont {C.~D.}\
  \bibnamefont {Panda}}, \bibinfo {author} {\bibfnamefont {A.~D.}\ \bibnamefont
  {West}}, \bibinfo {author} {\bibfnamefont {E.~P.}\ \bibnamefont {West}},\
  and\ \bibinfo {author} {\bibfnamefont {X.}~\bibnamefont {Wu}},\ }\bibfield
  {title} {\bibinfo {title} {{Improved limit on the electric dipole moment of
  the electron}},\ }\href {https://doi.org/10.1038/s41586-018-0599-8}
  {\bibfield  {journal} {\bibinfo  {journal} {Nature}\ }\textbf {\bibinfo
  {volume} {562}},\ \bibinfo {pages} {355} (\bibinfo {year}
  {2018})}\BibitemShut {NoStop}%
\bibitem [{\citenamefont {Roussy}\ \emph {et~al.}(2023)\citenamefont {Roussy},
  \citenamefont {Caldwell}, \citenamefont {Wright}, \citenamefont {Cairncross},
  \citenamefont {Shagam}, \citenamefont {Ng}, \citenamefont {Schlossberger},
  \citenamefont {Park}, \citenamefont {Wang}, \citenamefont {Ye},\ and\
  \citenamefont {Cornell}}]{Roussy2023}%
  \BibitemOpen
  \bibfield  {author} {\bibinfo {author} {\bibfnamefont {T.~S.}\ \bibnamefont
  {Roussy}}, \bibinfo {author} {\bibfnamefont {L.}~\bibnamefont {Caldwell}},
  \bibinfo {author} {\bibfnamefont {T.}~\bibnamefont {Wright}}, \bibinfo
  {author} {\bibfnamefont {W.~B.}\ \bibnamefont {Cairncross}}, \bibinfo
  {author} {\bibfnamefont {Y.}~\bibnamefont {Shagam}}, \bibinfo {author}
  {\bibfnamefont {K.~B.}\ \bibnamefont {Ng}}, \bibinfo {author} {\bibfnamefont
  {N.}~\bibnamefont {Schlossberger}}, \bibinfo {author} {\bibfnamefont {S.~Y.}\
  \bibnamefont {Park}}, \bibinfo {author} {\bibfnamefont {A.}~\bibnamefont
  {Wang}}, \bibinfo {author} {\bibfnamefont {J.}~\bibnamefont {Ye}},\ and\
  \bibinfo {author} {\bibfnamefont {E.~A.}\ \bibnamefont {Cornell}},\
  }\bibfield  {title} {\bibinfo {title} {An improved bound on the electron’s
  electric dipole moment},\ }\href {https://doi.org/10.1126/science.adg4084}
  {\bibfield  {journal} {\bibinfo  {journal} {Science}\ }\textbf {\bibinfo
  {volume} {381}},\ \bibinfo {pages} {46} (\bibinfo {year} {2023})}\BibitemShut
  {NoStop}%
\bibitem [{\citenamefont {Cairncross}\ and\ \citenamefont
  {Ye}(2019)}]{Cairncross2019}%
  \BibitemOpen
  \bibfield  {author} {\bibinfo {author} {\bibfnamefont {W.~B.}\ \bibnamefont
  {Cairncross}}\ and\ \bibinfo {author} {\bibfnamefont {J.}~\bibnamefont
  {Ye}},\ }\bibfield  {title} {\bibinfo {title} {{Atoms and molecules in the
  search for time-reversal symmetry violation}},\ }\href
  {https://doi.org/10.1038/s42254-019-0080-0} {\bibfield  {journal} {\bibinfo
  {journal} {Nature Reviews Physics}\ }\textbf {\bibinfo {volume} {1}},\
  \bibinfo {pages} {510} (\bibinfo {year} {2019})}\BibitemShut {NoStop}%
\bibitem [{\citenamefont {Arrowsmith-Kron}\ \emph {et~al.}(2024)\citenamefont
  {Arrowsmith-Kron}, \citenamefont {Athanasakis-Kaklamanakis}, \citenamefont
  {Au}, \citenamefont {Ballof}, \citenamefont {Berger}, \citenamefont
  {Borschevsky}, \citenamefont {Breier}, \citenamefont {Buchinger},
  \citenamefont {Budker}, \citenamefont {Caldwell}, \citenamefont {Charles},
  \citenamefont {Dattani}, \citenamefont {de~Groote}, \citenamefont {DeMille},
  \citenamefont {Dickel}, \citenamefont {Dobaczewski}, \citenamefont
  {Düllmann}, \citenamefont {Eliav}, \citenamefont {Engel}, \citenamefont
  {Fan}, \citenamefont {Flambaum}, \citenamefont {Flanagan}, \citenamefont
  {Gaiser}, \citenamefont {Ruiz}, \citenamefont {Gaul}, \citenamefont {Giesen},
  \citenamefont {Ginges}, \citenamefont {Gottberg}, \citenamefont {Gwinner},
  \citenamefont {Heinke}, \citenamefont {Hoekstra}, \citenamefont {Holt},
  \citenamefont {Hutzler}, \citenamefont {Jayich}, \citenamefont {Karthein},
  \citenamefont {Leach}, \citenamefont {Madison}, \citenamefont
  {Malbrunot-Ettenauer}, \citenamefont {Miyagi}, \citenamefont {Moore},
  \citenamefont {Moroch}, \citenamefont {Navratil}, \citenamefont {Nazarewicz},
  \citenamefont {Neyens}, \citenamefont {Norrgard}, \citenamefont {Nusgart},
  \citenamefont {Pašteka}, \citenamefont {Petrov}, \citenamefont {Plaß},
  \citenamefont {Ready}, \citenamefont {Reiter}, \citenamefont {Reponen},
  \citenamefont {Rothe}, \citenamefont {Safronova}, \citenamefont
  {Scheidenerger}, \citenamefont {Shindler}, \citenamefont {Singh},
  \citenamefont {Skripnikov}, \citenamefont {Titov}, \citenamefont {Udrescu},
  \citenamefont {Wilkins},\ and\ \citenamefont {Yang}}]{Arrowsmith-Kron2024}%
  \BibitemOpen
  \bibfield  {author} {\bibinfo {author} {\bibfnamefont {G.}~\bibnamefont
  {Arrowsmith-Kron}}, \bibinfo {author} {\bibfnamefont {M.}~\bibnamefont
  {Athanasakis-Kaklamanakis}}, \bibinfo {author} {\bibfnamefont
  {M.}~\bibnamefont {Au}}, \bibinfo {author} {\bibfnamefont {J.}~\bibnamefont
  {Ballof}}, \bibinfo {author} {\bibfnamefont {R.}~\bibnamefont {Berger}},
  \bibinfo {author} {\bibfnamefont {A.}~\bibnamefont {Borschevsky}}, \bibinfo
  {author} {\bibfnamefont {A.~A.}\ \bibnamefont {Breier}}, \bibinfo {author}
  {\bibfnamefont {F.}~\bibnamefont {Buchinger}}, \bibinfo {author}
  {\bibfnamefont {D.}~\bibnamefont {Budker}}, \bibinfo {author} {\bibfnamefont
  {L.}~\bibnamefont {Caldwell}}, \bibinfo {author} {\bibfnamefont
  {C.}~\bibnamefont {Charles}}, \bibinfo {author} {\bibfnamefont
  {N.}~\bibnamefont {Dattani}}, \bibinfo {author} {\bibfnamefont {R.~P.}\
  \bibnamefont {de~Groote}}, \bibinfo {author} {\bibfnamefont {D.}~\bibnamefont
  {DeMille}}, \bibinfo {author} {\bibfnamefont {T.}~\bibnamefont {Dickel}},
  \bibinfo {author} {\bibfnamefont {J.}~\bibnamefont {Dobaczewski}}, \bibinfo
  {author} {\bibfnamefont {C.~E.}\ \bibnamefont {Düllmann}}, \bibinfo {author}
  {\bibfnamefont {E.}~\bibnamefont {Eliav}}, \bibinfo {author} {\bibfnamefont
  {J.}~\bibnamefont {Engel}}, \bibinfo {author} {\bibfnamefont
  {M.}~\bibnamefont {Fan}}, \bibinfo {author} {\bibfnamefont {V.}~\bibnamefont
  {Flambaum}}, \bibinfo {author} {\bibfnamefont {K.~T.}\ \bibnamefont
  {Flanagan}}, \bibinfo {author} {\bibfnamefont {A.~N.}\ \bibnamefont
  {Gaiser}}, \bibinfo {author} {\bibfnamefont {R.~F.~G.}\ \bibnamefont {Ruiz}},
  \bibinfo {author} {\bibfnamefont {K.}~\bibnamefont {Gaul}}, \bibinfo {author}
  {\bibfnamefont {T.~F.}\ \bibnamefont {Giesen}}, \bibinfo {author}
  {\bibfnamefont {J.~S.~M.}\ \bibnamefont {Ginges}}, \bibinfo {author}
  {\bibfnamefont {A.}~\bibnamefont {Gottberg}}, \bibinfo {author}
  {\bibfnamefont {G.}~\bibnamefont {Gwinner}}, \bibinfo {author} {\bibfnamefont
  {R.}~\bibnamefont {Heinke}}, \bibinfo {author} {\bibfnamefont
  {S.}~\bibnamefont {Hoekstra}}, \bibinfo {author} {\bibfnamefont {J.~D.}\
  \bibnamefont {Holt}}, \bibinfo {author} {\bibfnamefont {N.~R.}\ \bibnamefont
  {Hutzler}}, \bibinfo {author} {\bibfnamefont {A.}~\bibnamefont {Jayich}},
  \bibinfo {author} {\bibfnamefont {J.}~\bibnamefont {Karthein}}, \bibinfo
  {author} {\bibfnamefont {K.~G.}\ \bibnamefont {Leach}}, \bibinfo {author}
  {\bibfnamefont {K.~W.}\ \bibnamefont {Madison}}, \bibinfo {author}
  {\bibfnamefont {S.}~\bibnamefont {Malbrunot-Ettenauer}}, \bibinfo {author}
  {\bibfnamefont {T.}~\bibnamefont {Miyagi}}, \bibinfo {author} {\bibfnamefont
  {I.~D.}\ \bibnamefont {Moore}}, \bibinfo {author} {\bibfnamefont
  {S.}~\bibnamefont {Moroch}}, \bibinfo {author} {\bibfnamefont
  {P.}~\bibnamefont {Navratil}}, \bibinfo {author} {\bibfnamefont
  {W.}~\bibnamefont {Nazarewicz}}, \bibinfo {author} {\bibfnamefont
  {G.}~\bibnamefont {Neyens}}, \bibinfo {author} {\bibfnamefont {E.~B.}\
  \bibnamefont {Norrgard}}, \bibinfo {author} {\bibfnamefont {N.}~\bibnamefont
  {Nusgart}}, \bibinfo {author} {\bibfnamefont {L.~F.}\ \bibnamefont
  {Pašteka}}, \bibinfo {author} {\bibfnamefont {A.~N.}\ \bibnamefont
  {Petrov}}, \bibinfo {author} {\bibfnamefont {W.~R.}\ \bibnamefont {Plaß}},
  \bibinfo {author} {\bibfnamefont {R.~A.}\ \bibnamefont {Ready}}, \bibinfo
  {author} {\bibfnamefont {M.~P.}\ \bibnamefont {Reiter}}, \bibinfo {author}
  {\bibfnamefont {M.}~\bibnamefont {Reponen}}, \bibinfo {author} {\bibfnamefont
  {S.}~\bibnamefont {Rothe}}, \bibinfo {author} {\bibfnamefont {M.~S.}\
  \bibnamefont {Safronova}}, \bibinfo {author} {\bibfnamefont {C.}~\bibnamefont
  {Scheidenerger}}, \bibinfo {author} {\bibfnamefont {A.}~\bibnamefont
  {Shindler}}, \bibinfo {author} {\bibfnamefont {J.~T.}\ \bibnamefont {Singh}},
  \bibinfo {author} {\bibfnamefont {L.~V.}\ \bibnamefont {Skripnikov}},
  \bibinfo {author} {\bibfnamefont {A.~V.}\ \bibnamefont {Titov}}, \bibinfo
  {author} {\bibfnamefont {S.-M.}\ \bibnamefont {Udrescu}}, \bibinfo {author}
  {\bibfnamefont {S.~G.}\ \bibnamefont {Wilkins}},\ and\ \bibinfo {author}
  {\bibfnamefont {X.}~\bibnamefont {Yang}},\ }\bibfield  {title} {\bibinfo
  {title} {Opportunities for fundamental physics research with radioactive
  molecules},\ }\href {https://doi.org/10.1088/1361-6633/ad1e39} {\bibfield
  {journal} {\bibinfo  {journal} {Rep. Prog. Phys.}\ }\textbf {\bibinfo
  {volume} {87}},\ \bibinfo {pages} {084301} (\bibinfo {year}
  {2024})}\BibitemShut {NoStop}%
\bibitem [{\citenamefont {Chupp}\ \emph {et~al.}(2019)\citenamefont {Chupp},
  \citenamefont {Fierlinger}, \citenamefont {Ramsey-Musolf},\ and\
  \citenamefont {Singh}}]{Chupp2019}%
  \BibitemOpen
  \bibfield  {author} {\bibinfo {author} {\bibfnamefont {T.~E.}\ \bibnamefont
  {Chupp}}, \bibinfo {author} {\bibfnamefont {P.}~\bibnamefont {Fierlinger}},
  \bibinfo {author} {\bibfnamefont {M.~J.}\ \bibnamefont {Ramsey-Musolf}},\
  and\ \bibinfo {author} {\bibfnamefont {J.~T.}\ \bibnamefont {Singh}},\
  }\bibfield  {title} {\bibinfo {title} {Electric dipole moments of atoms,
  molecules, nuclei, and particles},\ }\href
  {https://doi.org/10.1103/RevModPhys.91.015001} {\bibfield  {journal}
  {\bibinfo  {journal} {Rev. Mod. Phys.}\ }\textbf {\bibinfo {volume} {91}},\
  \bibinfo {pages} {015001} (\bibinfo {year} {2019})}\BibitemShut {NoStop}%
\bibitem [{\citenamefont {Sakharov}(1991)}]{Sakharov1991}%
  \BibitemOpen
  \bibfield  {author} {\bibinfo {author} {\bibfnamefont {A.~D.}\ \bibnamefont
  {Sakharov}},\ }\bibfield  {title} {\bibinfo {title} {Violation of {CP}
  invariance, {C} asymmetry, and baryon asymmetry of the universe},\ }\href
  {https://doi.org/10.1070/pu1991v034n05abeh002497} {\bibfield  {journal}
  {\bibinfo  {journal} {Phys.-Uspekhi}\ }\textbf {\bibinfo {volume} {34}},\
  \bibinfo {pages} {392} (\bibinfo {year} {1991})}\BibitemShut {NoStop}%
\bibitem [{\citenamefont {Dine}\ and\ \citenamefont
  {Kusenko}(2003)}]{Dine2003}%
  \BibitemOpen
  \bibfield  {author} {\bibinfo {author} {\bibfnamefont {M.}~\bibnamefont
  {Dine}}\ and\ \bibinfo {author} {\bibfnamefont {A.}~\bibnamefont {Kusenko}},\
  }\bibfield  {title} {\bibinfo {title} {Origin of the matter-antimatter
  asymmetry},\ }\href {https://doi.org/10.1103/RevModPhys.76.1} {\bibfield
  {journal} {\bibinfo  {journal} {Reviews of Modern Physics}\ }\textbf
  {\bibinfo {volume} {76}},\ \bibinfo {pages} {1} (\bibinfo {year} {2003})},\
  \bibinfo {note} {publisher: American Physical Society}\BibitemShut {NoStop}%
\bibitem [{\citenamefont {Fitch}\ \emph {et~al.}(2021)\citenamefont {Fitch},
  \citenamefont {Lim}, \citenamefont {Hinds}, \citenamefont {Sauer},\ and\
  \citenamefont {Tarbutt}}]{Fitch2020b}%
  \BibitemOpen
  \bibfield  {author} {\bibinfo {author} {\bibfnamefont {N.~J.}\ \bibnamefont
  {Fitch}}, \bibinfo {author} {\bibfnamefont {J.}~\bibnamefont {Lim}}, \bibinfo
  {author} {\bibfnamefont {E.~A.}\ \bibnamefont {Hinds}}, \bibinfo {author}
  {\bibfnamefont {B.~E.}\ \bibnamefont {Sauer}},\ and\ \bibinfo {author}
  {\bibfnamefont {M.~R.}\ \bibnamefont {Tarbutt}},\ }\bibfield  {title}
  {\bibinfo {title} {{Methods for measuring the electron's electric dipole
  moment using ultracold YbF molecules}},\ }\href
  {https://doi.org/10.1088/2058-9565/abc931} {\bibfield  {journal} {\bibinfo
  {journal} {Quantum Sci. Technol.}\ }\textbf {\bibinfo {volume} {6}},\
  \bibinfo {pages} {014006} (\bibinfo {year} {2021})}\BibitemShut {NoStop}%
\bibitem [{\citenamefont {Anderegg}\ \emph {et~al.}(2023)\citenamefont
  {Anderegg}, \citenamefont {Vilas}, \citenamefont {Hallas}, \citenamefont
  {Robichaud}, \citenamefont {Jadbabaie}, \citenamefont {Doyle},\ and\
  \citenamefont {Hutzler}}]{Anderegg2023}%
  \BibitemOpen
  \bibfield  {author} {\bibinfo {author} {\bibfnamefont {L.}~\bibnamefont
  {Anderegg}}, \bibinfo {author} {\bibfnamefont {N.~B.}\ \bibnamefont {Vilas}},
  \bibinfo {author} {\bibfnamefont {C.}~\bibnamefont {Hallas}}, \bibinfo
  {author} {\bibfnamefont {P.}~\bibnamefont {Robichaud}}, \bibinfo {author}
  {\bibfnamefont {A.}~\bibnamefont {Jadbabaie}}, \bibinfo {author}
  {\bibfnamefont {J.~M.}\ \bibnamefont {Doyle}},\ and\ \bibinfo {author}
  {\bibfnamefont {N.~R.}\ \bibnamefont {Hutzler}},\ }\bibfield  {title}
  {\bibinfo {title} {{Quantum Control of Trapped Polyatomic Molecules for eEDM
  Searches}},\ }\href {https://doi.org/10.1126/science.adg8155} {\bibfield
  {journal} {\bibinfo  {journal} {Science}\ }\textbf {\bibinfo {volume}
  {382}},\ \bibinfo {pages} {665} (\bibinfo {year} {2023})}\BibitemShut
  {NoStop}%
\bibitem [{\citenamefont {Bause}\ \emph {et~al.}(2024)\citenamefont {Bause},
  \citenamefont {Balasubramanian}, \citenamefont {Fikkers}, \citenamefont
  {Prinsen}, \citenamefont {Steinebach}, \citenamefont {Jadbabaie},
  \citenamefont {Hutzler}, \citenamefont {Aucar}, \citenamefont {Pašteka},
  \citenamefont {Borschevsky},\ and\ \citenamefont
  {Hoekstra}}]{Bause2024arxiv}%
  \BibitemOpen
  \bibfield  {author} {\bibinfo {author} {\bibfnamefont {R.}~\bibnamefont
  {Bause}}, \bibinfo {author} {\bibfnamefont {N.}~\bibnamefont
  {Balasubramanian}}, \bibinfo {author} {\bibfnamefont {T.}~\bibnamefont
  {Fikkers}}, \bibinfo {author} {\bibfnamefont {E.~H.}\ \bibnamefont
  {Prinsen}}, \bibinfo {author} {\bibfnamefont {K.}~\bibnamefont {Steinebach}},
  \bibinfo {author} {\bibfnamefont {A.}~\bibnamefont {Jadbabaie}}, \bibinfo
  {author} {\bibfnamefont {N.~R.}\ \bibnamefont {Hutzler}}, \bibinfo {author}
  {\bibfnamefont {I.~A.}\ \bibnamefont {Aucar}}, \bibinfo {author}
  {\bibfnamefont {L.~F.}\ \bibnamefont {Pašteka}}, \bibinfo {author}
  {\bibfnamefont {A.}~\bibnamefont {Borschevsky}},\ and\ \bibinfo {author}
  {\bibfnamefont {S.}~\bibnamefont {Hoekstra}},\ }\href
  {https://arxiv.org/abs/2411.00441} {\bibinfo {title} {Prospects for measuring
  the electron's electric dipole moment with polyatomic molecules in an optical
  lattice}} (\bibinfo {year} {2024}),\ \Eprint
  {https://arxiv.org/abs/2411.00441} {arXiv:2411.00441 [physics.atom-ph]}
  \BibitemShut {NoStop}%
\bibitem [{\citenamefont {Zeng}\ \emph {et~al.}(2024)\citenamefont {Zeng},
  \citenamefont {Deng}, \citenamefont {Yang},\ and\ \citenamefont
  {Yan}}]{Zeng2024}%
  \BibitemOpen
  \bibfield  {author} {\bibinfo {author} {\bibfnamefont {Z.}~\bibnamefont
  {Zeng}}, \bibinfo {author} {\bibfnamefont {S.}~\bibnamefont {Deng}}, \bibinfo
  {author} {\bibfnamefont {S.}~\bibnamefont {Yang}},\ and\ \bibinfo {author}
  {\bibfnamefont {B.}~\bibnamefont {Yan}},\ }\bibfield  {title} {\bibinfo
  {title} {Three-dimensional magneto-optical trapping of barium monofluoride},\
  }\href {https://doi.org/10.1103/PhysRevLett.133.143404} {\bibfield  {journal}
  {\bibinfo  {journal} {Phys. Rev. Lett.}\ }\textbf {\bibinfo {volume} {133}},\
  \bibinfo {pages} {143404} (\bibinfo {year} {2024})}\BibitemShut {NoStop}%
\bibitem [{\citenamefont {Kozlov}(1997)}]{Kozlov1997}%
  \BibitemOpen
  \bibfield  {author} {\bibinfo {author} {\bibfnamefont {M.~G.}\ \bibnamefont
  {Kozlov}},\ }\bibfield  {title} {\bibinfo {title} {{Enhancement of the
  electric dipole moment of the electron in the YbF molecule}},\ }\href@noop {}
  {\bibfield  {journal} {\bibinfo  {journal} {J. Phys. B}\ }\textbf {\bibinfo
  {volume} {30}},\ \bibinfo {pages} {L607} (\bibinfo {year}
  {1997})}\BibitemShut {NoStop}%
\bibitem [{\citenamefont {Hudson}\ \emph {et~al.}(2002)\citenamefont {Hudson},
  \citenamefont {Sauer}, \citenamefont {Tarbutt},\ and\ \citenamefont
  {Hinds}}]{Hudson2002}%
  \BibitemOpen
  \bibfield  {author} {\bibinfo {author} {\bibfnamefont {J.~J.}\ \bibnamefont
  {Hudson}}, \bibinfo {author} {\bibfnamefont {B.~E.}\ \bibnamefont {Sauer}},
  \bibinfo {author} {\bibfnamefont {M.~R.}\ \bibnamefont {Tarbutt}},\ and\
  \bibinfo {author} {\bibfnamefont {E.~A.}\ \bibnamefont {Hinds}},\ }\bibfield
  {title} {\bibinfo {title} {Measurement of the electron electric dipole moment
  using {YbF} molecules},\ }\href@noop {} {\bibfield  {journal} {\bibinfo
  {journal} {Phys. Rev. Lett.}\ }\textbf {\bibinfo {volume} {89}},\ \bibinfo
  {pages} {023003} (\bibinfo {year} {2002})}\BibitemShut {NoStop}%
\bibitem [{\citenamefont {Tarbutt}\ \emph {et~al.}(2013)\citenamefont
  {Tarbutt}, \citenamefont {Sauer}, \citenamefont {Hudson},\ and\ \citenamefont
  {Hinds}}]{Tarbutt2013}%
  \BibitemOpen
  \bibfield  {author} {\bibinfo {author} {\bibfnamefont {M.~R.}\ \bibnamefont
  {Tarbutt}}, \bibinfo {author} {\bibfnamefont {B.~E.}\ \bibnamefont {Sauer}},
  \bibinfo {author} {\bibfnamefont {J.~J.}\ \bibnamefont {Hudson}},\ and\
  \bibinfo {author} {\bibfnamefont {E.~A.}\ \bibnamefont {Hinds}},\ }\bibfield
  {title} {\bibinfo {title} {{Design for a fountain of YbF molecules to measure
  the electron's electric dipole moment}},\ }\href
  {https://doi.org/10.1088/1367-2630/15/5/053034} {\bibfield  {journal}
  {\bibinfo  {journal} {New J. Phys.}\ }\textbf {\bibinfo {volume} {15}},\
  \bibinfo {pages} {053034} (\bibinfo {year} {2013})}\BibitemShut {NoStop}%
\bibitem [{\citenamefont {Ho}\ \emph {et~al.}(2023)\citenamefont {Ho},
  \citenamefont {Lim}, \citenamefont {Sauer},\ and\ \citenamefont
  {Tarbutt}}]{Ho2023}%
  \BibitemOpen
  \bibfield  {author} {\bibinfo {author} {\bibfnamefont {C.~J.}\ \bibnamefont
  {Ho}}, \bibinfo {author} {\bibfnamefont {J.}~\bibnamefont {Lim}}, \bibinfo
  {author} {\bibfnamefont {B.~E.}\ \bibnamefont {Sauer}},\ and\ \bibinfo
  {author} {\bibfnamefont {M.~R.}\ \bibnamefont {Tarbutt}},\ }\bibfield
  {title} {\bibinfo {title} {Measuring the nuclear magnetic quadrupole moment
  in heavy polar molecules},\ }\href
  {https://doi.org/10.3389/fphy.2023.1086980} {\bibfield  {journal} {\bibinfo
  {journal} {Front. Phys.}\ }\textbf {\bibinfo {volume} {11}},\ \bibinfo
  {pages} {1086980} (\bibinfo {year} {2023})}\BibitemShut {NoStop}%
\bibitem [{\citenamefont {Zheng}\ \emph {et~al.}(2022)\citenamefont {Zheng},
  \citenamefont {Yang}, \citenamefont {Wang}, \citenamefont {Singh},
  \citenamefont {Xiong}, \citenamefont {Xia},\ and\ \citenamefont
  {Lu}}]{Zheng2022}%
  \BibitemOpen
  \bibfield  {author} {\bibinfo {author} {\bibfnamefont {T.~A.}\ \bibnamefont
  {Zheng}}, \bibinfo {author} {\bibfnamefont {Y.~A.}\ \bibnamefont {Yang}},
  \bibinfo {author} {\bibfnamefont {S.-Z.}\ \bibnamefont {Wang}}, \bibinfo
  {author} {\bibfnamefont {J.~T.}\ \bibnamefont {Singh}}, \bibinfo {author}
  {\bibfnamefont {Z.-X.}\ \bibnamefont {Xiong}}, \bibinfo {author}
  {\bibfnamefont {T.}~\bibnamefont {Xia}},\ and\ \bibinfo {author}
  {\bibfnamefont {Z.-T.}\ \bibnamefont {Lu}},\ }\bibfield  {title} {\bibinfo
  {title} {Measurement of the electric dipole moment of $^{171}\mathrm{Yb}$
  atoms in an optical dipole trap},\ }\href
  {https://doi.org/10.1103/PhysRevLett.129.083001} {\bibfield  {journal}
  {\bibinfo  {journal} {Phys. Rev. Lett.}\ }\textbf {\bibinfo {volume} {129}},\
  \bibinfo {pages} {083001} (\bibinfo {year} {2022})}\BibitemShut {NoStop}%
\bibitem [{\citenamefont {Lim}\ \emph {et~al.}(2018)\citenamefont {Lim},
  \citenamefont {Almond}, \citenamefont {Trigatzis}, \citenamefont {Devlin},
  \citenamefont {Fitch}, \citenamefont {Sauer}, \citenamefont {Tarbutt},\ and\
  \citenamefont {Hinds}}]{Lim2018}%
  \BibitemOpen
  \bibfield  {author} {\bibinfo {author} {\bibfnamefont {J.}~\bibnamefont
  {Lim}}, \bibinfo {author} {\bibfnamefont {J.~R.}\ \bibnamefont {Almond}},
  \bibinfo {author} {\bibfnamefont {M.~A.}\ \bibnamefont {Trigatzis}}, \bibinfo
  {author} {\bibfnamefont {J.~A.}\ \bibnamefont {Devlin}}, \bibinfo {author}
  {\bibfnamefont {N.~J.}\ \bibnamefont {Fitch}}, \bibinfo {author}
  {\bibfnamefont {B.~E.}\ \bibnamefont {Sauer}}, \bibinfo {author}
  {\bibfnamefont {M.~R.}\ \bibnamefont {Tarbutt}},\ and\ \bibinfo {author}
  {\bibfnamefont {E.~A.}\ \bibnamefont {Hinds}},\ }\bibfield  {title} {\bibinfo
  {title} {Laser cooled {YbF} molecules for measuring the electron's electric
  dipole moment},\ }\href {https://doi.org/10.1103/PhysRevLett.120.123201}
  {\bibfield  {journal} {\bibinfo  {journal} {Phys. Rev. Lett.}\ }\textbf
  {\bibinfo {volume} {120}},\ \bibinfo {pages} {123201} (\bibinfo {year}
  {2018})}\BibitemShut {NoStop}%
\bibitem [{\citenamefont {Alauze}\ \emph {et~al.}(2021)\citenamefont {Alauze},
  \citenamefont {Lim}, \citenamefont {Trigatzis}, \citenamefont {Swarbrick},
  \citenamefont {Collings}, \citenamefont {Fitch}, \citenamefont {Sauer},\ and\
  \citenamefont {Tarbutt}}]{Alauze2021}%
  \BibitemOpen
  \bibfield  {author} {\bibinfo {author} {\bibfnamefont {X.}~\bibnamefont
  {Alauze}}, \bibinfo {author} {\bibfnamefont {J.}~\bibnamefont {Lim}},
  \bibinfo {author} {\bibfnamefont {M.~A.}\ \bibnamefont {Trigatzis}}, \bibinfo
  {author} {\bibfnamefont {S.}~\bibnamefont {Swarbrick}}, \bibinfo {author}
  {\bibfnamefont {F.~J.}\ \bibnamefont {Collings}}, \bibinfo {author}
  {\bibfnamefont {N.~J.}\ \bibnamefont {Fitch}}, \bibinfo {author}
  {\bibfnamefont {B.~E.}\ \bibnamefont {Sauer}},\ and\ \bibinfo {author}
  {\bibfnamefont {M.~R.}\ \bibnamefont {Tarbutt}},\ }\bibfield  {title}
  {\bibinfo {title} {{An ultracold molecular beam for testing fundamental
  physics}},\ }\href {https://doi.org/https://doi.org/10.1088/2058-9565/ac107e}
  {\bibfield  {journal} {\bibinfo  {journal} {Quantum Sci. Technol.}\ }\textbf
  {\bibinfo {volume} {6}},\ \bibinfo {pages} {044005} (\bibinfo {year}
  {2021})}\BibitemShut {NoStop}%
\bibitem [{\citenamefont {Barry}\ \emph {et~al.}(2012)\citenamefont {Barry},
  \citenamefont {Shuman}, \citenamefont {Norrgard},\ and\ \citenamefont
  {DeMille}}]{Barry2012}%
  \BibitemOpen
  \bibfield  {author} {\bibinfo {author} {\bibfnamefont {J.~F.}\ \bibnamefont
  {Barry}}, \bibinfo {author} {\bibfnamefont {E.~S.}\ \bibnamefont {Shuman}},
  \bibinfo {author} {\bibfnamefont {E.~B.}\ \bibnamefont {Norrgard}},\ and\
  \bibinfo {author} {\bibfnamefont {D.}~\bibnamefont {DeMille}},\ }\bibfield
  {title} {\bibinfo {title} {{Laser radiation pressure slowing of a molecular
  beam}},\ }\href {https://doi.org/10.1103/PhysRevLett.108.103002} {\bibfield
  {journal} {\bibinfo  {journal} {Phys. Rev. Lett.}\ }\textbf {\bibinfo
  {volume} {108}},\ \bibinfo {pages} {103002} (\bibinfo {year}
  {2012})}\BibitemShut {NoStop}%
\bibitem [{\citenamefont {Zhelyazkova}\ \emph {et~al.}(2014)\citenamefont
  {Zhelyazkova}, \citenamefont {Cournol}, \citenamefont {Wall}, \citenamefont
  {Matsushima}, \citenamefont {Hudson}, \citenamefont {Hinds}, \citenamefont
  {Tarbutt},\ and\ \citenamefont {Sauer}}]{Zhelyazkova2014}%
  \BibitemOpen
  \bibfield  {author} {\bibinfo {author} {\bibfnamefont {V.}~\bibnamefont
  {Zhelyazkova}}, \bibinfo {author} {\bibfnamefont {A.}~\bibnamefont
  {Cournol}}, \bibinfo {author} {\bibfnamefont {T.~E.}\ \bibnamefont {Wall}},
  \bibinfo {author} {\bibfnamefont {A.}~\bibnamefont {Matsushima}}, \bibinfo
  {author} {\bibfnamefont {J.~J.}\ \bibnamefont {Hudson}}, \bibinfo {author}
  {\bibfnamefont {E.~A.}\ \bibnamefont {Hinds}}, \bibinfo {author}
  {\bibfnamefont {M.~R.}\ \bibnamefont {Tarbutt}},\ and\ \bibinfo {author}
  {\bibfnamefont {B.~E.}\ \bibnamefont {Sauer}},\ }\bibfield  {title} {\bibinfo
  {title} {{Laser cooling and slowing of CaF molecules}},\ }\href
  {https://doi.org/10.1103/PhysRevA.89.053416} {\bibfield  {journal} {\bibinfo
  {journal} {Phys. Rev. A}\ }\textbf {\bibinfo {volume} {89}},\ \bibinfo
  {pages} {053416} (\bibinfo {year} {2014})}\BibitemShut {NoStop}%
\bibitem [{\citenamefont {Truppe}\ \emph {et~al.}(2017)\citenamefont {Truppe},
  \citenamefont {Williams}, \citenamefont {Fitch}, \citenamefont {Hambach},
  \citenamefont {Wall}, \citenamefont {Hinds}, \citenamefont {Sauer},\ and\
  \citenamefont {Tarbutt}}]{Truppe2017}%
  \BibitemOpen
  \bibfield  {author} {\bibinfo {author} {\bibfnamefont {S.}~\bibnamefont
  {Truppe}}, \bibinfo {author} {\bibfnamefont {H.~J.}\ \bibnamefont
  {Williams}}, \bibinfo {author} {\bibfnamefont {N.~J.}\ \bibnamefont {Fitch}},
  \bibinfo {author} {\bibfnamefont {M.}~\bibnamefont {Hambach}}, \bibinfo
  {author} {\bibfnamefont {T.~E.}\ \bibnamefont {Wall}}, \bibinfo {author}
  {\bibfnamefont {E.~A.}\ \bibnamefont {Hinds}}, \bibinfo {author}
  {\bibfnamefont {B.~E.}\ \bibnamefont {Sauer}},\ and\ \bibinfo {author}
  {\bibfnamefont {M.~R.}\ \bibnamefont {Tarbutt}},\ }\bibfield  {title}
  {\bibinfo {title} {An intense, cold, velocity-controlled molecular beam by
  frequency-chirped laser slowing},\ }\href
  {https://doi.org/10.1088/1367-2630/aa5ca2} {\bibfield  {journal} {\bibinfo
  {journal} {New J. Phys.}\ }\textbf {\bibinfo {volume} {19}},\ \bibinfo
  {pages} {022001} (\bibinfo {year} {2017})}\BibitemShut {NoStop}%
\bibitem [{\citenamefont {Hemmerling}\ \emph {et~al.}(2016)\citenamefont
  {Hemmerling}, \citenamefont {Chae}, \citenamefont {Ravi}, \citenamefont
  {Anderegg}, \citenamefont {Drayna}, \citenamefont {Hutzler}, \citenamefont
  {Collopy}, \citenamefont {Ye}, \citenamefont {Ketterle},\ and\ \citenamefont
  {Doyle}}]{Hemmerling2016}%
  \BibitemOpen
  \bibfield  {author} {\bibinfo {author} {\bibfnamefont {B.}~\bibnamefont
  {Hemmerling}}, \bibinfo {author} {\bibfnamefont {E.}~\bibnamefont {Chae}},
  \bibinfo {author} {\bibfnamefont {A.}~\bibnamefont {Ravi}}, \bibinfo {author}
  {\bibfnamefont {L.}~\bibnamefont {Anderegg}}, \bibinfo {author}
  {\bibfnamefont {G.~K.}\ \bibnamefont {Drayna}}, \bibinfo {author}
  {\bibfnamefont {N.~R.}\ \bibnamefont {Hutzler}}, \bibinfo {author}
  {\bibfnamefont {A.~L.}\ \bibnamefont {Collopy}}, \bibinfo {author}
  {\bibfnamefont {J.}~\bibnamefont {Ye}}, \bibinfo {author} {\bibfnamefont
  {W.}~\bibnamefont {Ketterle}},\ and\ \bibinfo {author} {\bibfnamefont
  {J.~M.}\ \bibnamefont {Doyle}},\ }\bibfield  {title} {\bibinfo {title}
  {{Laser slowing of CaF molecules to near the capture velocity of a molecular
  MOT}},\ }\href@noop {} {\bibfield  {journal} {\bibinfo  {journal} {J. Phys.
  B}\ }\textbf {\bibinfo {volume} {49}},\ \bibinfo {pages} {174001} (\bibinfo
  {year} {2016})}\BibitemShut {NoStop}%
\bibitem [{\citenamefont {Zhang}\ \emph {et~al.}(2022)\citenamefont {Zhang},
  \citenamefont {Zhang}, \citenamefont {Cheng}, \citenamefont {Steimle},\ and\
  \citenamefont {Tarbutt}}]{Zhang2022}%
  \BibitemOpen
  \bibfield  {author} {\bibinfo {author} {\bibfnamefont {C.}~\bibnamefont
  {Zhang}}, \bibinfo {author} {\bibfnamefont {C.}~\bibnamefont {Zhang}},
  \bibinfo {author} {\bibfnamefont {L.}~\bibnamefont {Cheng}}, \bibinfo
  {author} {\bibfnamefont {T.~C.}\ \bibnamefont {Steimle}},\ and\ \bibinfo
  {author} {\bibfnamefont {M.~R.}\ \bibnamefont {Tarbutt}},\ }\bibfield
  {title} {\bibinfo {title} {{Inner-shell excitation in the YbF molecule and
  its impact on laser cooling}},\ }\href
  {https://doi.org/10.1016/j.jms.2022.111625} {\bibfield  {journal} {\bibinfo
  {journal} {J. Mol. Spectrosc.}\ }\textbf {\bibinfo {volume} {386}},\ \bibinfo
  {pages} {111625} (\bibinfo {year} {2022})}\BibitemShut {NoStop}%
\bibitem [{\citenamefont {Popa}\ \emph {et~al.}(2024)\citenamefont {Popa},
  \citenamefont {Schaller}, \citenamefont {Fielicke}, \citenamefont {Lim},
  \citenamefont {Sartakov}, \citenamefont {Tarbutt},\ and\ \citenamefont
  {Meijer}}]{Popa2024}%
  \BibitemOpen
  \bibfield  {author} {\bibinfo {author} {\bibfnamefont {S.}~\bibnamefont
  {Popa}}, \bibinfo {author} {\bibfnamefont {S.}~\bibnamefont {Schaller}},
  \bibinfo {author} {\bibfnamefont {A.}~\bibnamefont {Fielicke}}, \bibinfo
  {author} {\bibfnamefont {J.}~\bibnamefont {Lim}}, \bibinfo {author}
  {\bibfnamefont {B.~G.}\ \bibnamefont {Sartakov}}, \bibinfo {author}
  {\bibfnamefont {M.~R.}\ \bibnamefont {Tarbutt}},\ and\ \bibinfo {author}
  {\bibfnamefont {G.}~\bibnamefont {Meijer}},\ }\bibfield  {title} {\bibinfo
  {title} {{Understanding Inner-Shell Excitations in Molecules through
  Spectroscopy of the 4f Hole States of YbF}},\ }\href
  {https://doi.org/10.1103/PhysRevX.14.021035} {\bibfield  {journal} {\bibinfo
  {journal} {Phys. Rev. X}\ }\textbf {\bibinfo {volume} {14}},\ \bibinfo
  {pages} {021035} (\bibinfo {year} {2024})}\BibitemShut {NoStop}%
\bibitem [{\citenamefont {White}\ \emph {et~al.}(2024)\citenamefont {White},
  \citenamefont {Popa}, \citenamefont {Mellado-Mu\~noz}, \citenamefont {Fitch},
  \citenamefont {Sauer}, \citenamefont {Lim},\ and\ \citenamefont
  {Tarbutt}}]{White2024}%
  \BibitemOpen
  \bibfield  {author} {\bibinfo {author} {\bibfnamefont {A.~D.}\ \bibnamefont
  {White}}, \bibinfo {author} {\bibfnamefont {S.}~\bibnamefont {Popa}},
  \bibinfo {author} {\bibfnamefont {J.}~\bibnamefont {Mellado-Mu\~noz}},
  \bibinfo {author} {\bibfnamefont {N.~J.}\ \bibnamefont {Fitch}}, \bibinfo
  {author} {\bibfnamefont {B.~E.}\ \bibnamefont {Sauer}}, \bibinfo {author}
  {\bibfnamefont {J.}~\bibnamefont {Lim}},\ and\ \bibinfo {author}
  {\bibfnamefont {M.~R.}\ \bibnamefont {Tarbutt}},\ }\bibfield  {title}
  {\bibinfo {title} {Slow molecular beams from a cryogenic buffer gas source},\
  }\href {https://doi.org/10.1103/PhysRevResearch.6.043232} {\bibfield
  {journal} {\bibinfo  {journal} {Phys. Rev. Res.}\ }\textbf {\bibinfo {volume}
  {6}},\ \bibinfo {pages} {043232} (\bibinfo {year} {2024})}\BibitemShut
  {NoStop}%
\bibitem [{Note1()}]{Note1}%
  \BibitemOpen
  \bibinfo {note} {These labels specify the energy above the ground state, in
  THz. The same states are sometimes also called [18.58] and [18.71]~\cite
  {Popa2024} where the label is the energy in thousands of
  cm$^{-1}$.}\BibitemShut {Stop}%
\bibitem [{\citenamefont {Lim}\ \emph {et~al.}(2017)\citenamefont {Lim},
  \citenamefont {Almond}, \citenamefont {Tarbutt}, \citenamefont {Nguyen},\
  and\ \citenamefont {Steimle}}]{Lim2017}%
  \BibitemOpen
  \bibfield  {author} {\bibinfo {author} {\bibfnamefont {J.}~\bibnamefont
  {Lim}}, \bibinfo {author} {\bibfnamefont {J.~R.}\ \bibnamefont {Almond}},
  \bibinfo {author} {\bibfnamefont {M.~R.}\ \bibnamefont {Tarbutt}}, \bibinfo
  {author} {\bibfnamefont {D.~T.}\ \bibnamefont {Nguyen}},\ and\ \bibinfo
  {author} {\bibfnamefont {T.~C.}\ \bibnamefont {Steimle}},\ }\bibfield
  {title} {\bibinfo {title} {{The [557]-X$^2\Sigma^{+}$ and
  [561]-X$^2\Sigma^{+}$ bands of ytterbium fluoride, $^{174}$YbF}},\ }\href
  {https://doi.org/10.1016/j.jms.2017.06.007} {\bibfield  {journal} {\bibinfo
  {journal} {J. Mol. Spectrosc.}\ }\textbf {\bibinfo {volume} {338}},\ \bibinfo
  {pages} {81} (\bibinfo {year} {2017})}\BibitemShut {NoStop}%
\bibitem [{\citenamefont {Zhuang}\ \emph {et~al.}(2011)\citenamefont {Zhuang},
  \citenamefont {Le}, \citenamefont {Steimle}, \citenamefont {Bulleid},
  \citenamefont {Smallman}, \citenamefont {Hendricks}, \citenamefont {Skoff},
  \citenamefont {Hudson}, \citenamefont {Sauer}, \citenamefont {Hinds},\ and\
  \citenamefont {Tarbutt}}]{Zhuang2011}%
  \BibitemOpen
  \bibfield  {author} {\bibinfo {author} {\bibfnamefont {X.}~\bibnamefont
  {Zhuang}}, \bibinfo {author} {\bibfnamefont {A.}~\bibnamefont {Le}}, \bibinfo
  {author} {\bibfnamefont {T.~C.}\ \bibnamefont {Steimle}}, \bibinfo {author}
  {\bibfnamefont {N.~E.}\ \bibnamefont {Bulleid}}, \bibinfo {author}
  {\bibfnamefont {I.~J.}\ \bibnamefont {Smallman}}, \bibinfo {author}
  {\bibfnamefont {R.~J.}\ \bibnamefont {Hendricks}}, \bibinfo {author}
  {\bibfnamefont {S.~M.}\ \bibnamefont {Skoff}}, \bibinfo {author}
  {\bibfnamefont {J.~J.}\ \bibnamefont {Hudson}}, \bibinfo {author}
  {\bibfnamefont {B.~E.}\ \bibnamefont {Sauer}}, \bibinfo {author}
  {\bibfnamefont {E.~A.}\ \bibnamefont {Hinds}},\ and\ \bibinfo {author}
  {\bibfnamefont {M.~R.}\ \bibnamefont {Tarbutt}},\ }\bibfield  {title}
  {\bibinfo {title} {{Franck-Condon factors and radiative lifetime of the
  A$^{2}\Pi_{1/2}$-X$^{2}\Sigma^{+}$ transition of ytterbium monofluoride,
  YbF}},\ }\href@noop {} {\bibfield  {journal} {\bibinfo  {journal} {Phys.
  Chem. Chem. Phys.}\ }\textbf {\bibinfo {volume} {13}},\ \bibinfo {pages}
  {19013} (\bibinfo {year} {2011})}\BibitemShut {NoStop}%
\bibitem [{\citenamefont {Berkeland}\ and\ \citenamefont
  {Boshier}(2002)}]{Berkeland2002}%
  \BibitemOpen
  \bibfield  {author} {\bibinfo {author} {\bibfnamefont {D.~J.}\ \bibnamefont
  {Berkeland}}\ and\ \bibinfo {author} {\bibfnamefont {M.~G.}\ \bibnamefont
  {Boshier}},\ }\bibfield  {title} {\bibinfo {title} {{Destabilization of dark
  states and optical spectroscopy in Zeeman-degenerate atomic systems}},\
  }\href {https://doi.org/10.1103/PhysRevA.65.033413} {\bibfield  {journal}
  {\bibinfo  {journal} {Phys. Rev. A}\ }\textbf {\bibinfo {volume} {65}},\
  \bibinfo {pages} {033413} (\bibinfo {year} {2002})}\BibitemShut {NoStop}%
\bibitem [{\citenamefont {Fitch}\ and\ \citenamefont
  {Tarbutt}(2021)}]{Fitch2021b}%
  \BibitemOpen
  \bibfield  {author} {\bibinfo {author} {\bibfnamefont {N.~J.}\ \bibnamefont
  {Fitch}}\ and\ \bibinfo {author} {\bibfnamefont {M.~R.}\ \bibnamefont
  {Tarbutt}},\ }\bibfield  {title} {\bibinfo {title} {Laser-cooled molecules},\
  }in\ \href {https://doi.org/10.1016/bs.aamop.2021.04.003} {\emph {\bibinfo
  {booktitle} {Adv. At. Mol. Opt. Phys.}}},\ Vol.~\bibinfo {volume} {70}\
  (\bibinfo  {publisher} {Elsevier},\ \bibinfo {year} {2021})\ pp.\ \bibinfo
  {pages} {157--262}\BibitemShut {NoStop}%
\bibitem [{\citenamefont {Collopy}\ \emph {et~al.}(2018)\citenamefont
  {Collopy}, \citenamefont {Ding}, \citenamefont {Wu}, \citenamefont
  {Finneran}, \citenamefont {Anderegg}, \citenamefont {Augenbraun},
  \citenamefont {Doyle},\ and\ \citenamefont {Ye}}]{Collopy2018}%
  \BibitemOpen
  \bibfield  {author} {\bibinfo {author} {\bibfnamefont {A.~L.}\ \bibnamefont
  {Collopy}}, \bibinfo {author} {\bibfnamefont {S.}~\bibnamefont {Ding}},
  \bibinfo {author} {\bibfnamefont {Y.}~\bibnamefont {Wu}}, \bibinfo {author}
  {\bibfnamefont {I.~A.}\ \bibnamefont {Finneran}}, \bibinfo {author}
  {\bibfnamefont {L.}~\bibnamefont {Anderegg}}, \bibinfo {author}
  {\bibfnamefont {B.~L.}\ \bibnamefont {Augenbraun}}, \bibinfo {author}
  {\bibfnamefont {J.~M.}\ \bibnamefont {Doyle}},\ and\ \bibinfo {author}
  {\bibfnamefont {J.}~\bibnamefont {Ye}},\ }\bibfield  {title} {\bibinfo
  {title} {{3D magneto-optical trap of yttrium monoxide}},\ }\href
  {https://doi.org/10.1103/PhysRevLett.121.213201} {\bibfield  {journal}
  {\bibinfo  {journal} {Phys. Rev. Lett.}\ }\textbf {\bibinfo {volume} {121}},\
  \bibinfo {pages} {213201} (\bibinfo {year} {2018})}\BibitemShut {NoStop}%
\bibitem [{\citenamefont {Langin}\ and\ \citenamefont
  {DeMille}(2023)}]{Langin2023}%
  \BibitemOpen
  \bibfield  {author} {\bibinfo {author} {\bibfnamefont {T.~K.}\ \bibnamefont
  {Langin}}\ and\ \bibinfo {author} {\bibfnamefont {D.}~\bibnamefont
  {DeMille}},\ }\bibfield  {title} {\bibinfo {title} {Toward improved loading,
  cooling, and trapping of molecules in magneto-optical traps},\ }\href
  {https://doi.org/10.1088/1367-2630/acc34d} {\bibfield  {journal} {\bibinfo
  {journal} {New J. Phys.}\ }\textbf {\bibinfo {volume} {25}},\ \bibinfo
  {pages} {043005} (\bibinfo {year} {2023})}\BibitemShut {NoStop}%
\bibitem [{\citenamefont {Petzold}\ \emph {et~al.}(2013)\citenamefont
  {Petzold}, \citenamefont {Kaebert}, \citenamefont {Gersema}, \citenamefont
  {Siercke},\ and\ \citenamefont {Ospelkaus}}]{Petzold2018}%
  \BibitemOpen
  \bibfield  {author} {\bibinfo {author} {\bibfnamefont {M.}~\bibnamefont
  {Petzold}}, \bibinfo {author} {\bibfnamefont {P.}~\bibnamefont {Kaebert}},
  \bibinfo {author} {\bibfnamefont {P.}~\bibnamefont {Gersema}}, \bibinfo
  {author} {\bibfnamefont {M.}~\bibnamefont {Siercke}},\ and\ \bibinfo {author}
  {\bibfnamefont {S.}~\bibnamefont {Ospelkaus}},\ }\bibfield  {title} {\bibinfo
  {title} {{A Zeeman slower for diatomic molecules}},\ }\href
  {https://doi.org/10.1088/1367-2630/aab9f5} {\bibfield  {journal} {\bibinfo
  {journal} {New J. Phys.}\ }\textbf {\bibinfo {volume} {20}},\ \bibinfo
  {pages} {042001} (\bibinfo {year} {2013})}\BibitemShut {NoStop}%
\bibitem [{\citenamefont {Kaebert}\ \emph {et~al.}(2021)\citenamefont
  {Kaebert}, \citenamefont {Stepanova}, \citenamefont {Poll}, \citenamefont
  {Petzold}, \citenamefont {Xu}, \citenamefont {Siercke},\ and\ \citenamefont
  {Ospelkaus}}]{Kaebert2021}%
  \BibitemOpen
  \bibfield  {author} {\bibinfo {author} {\bibfnamefont {P.}~\bibnamefont
  {Kaebert}}, \bibinfo {author} {\bibfnamefont {M.}~\bibnamefont {Stepanova}},
  \bibinfo {author} {\bibfnamefont {T.}~\bibnamefont {Poll}}, \bibinfo {author}
  {\bibfnamefont {M.}~\bibnamefont {Petzold}}, \bibinfo {author} {\bibfnamefont
  {S.}~\bibnamefont {Xu}}, \bibinfo {author} {\bibfnamefont {M.}~\bibnamefont
  {Siercke}},\ and\ \bibinfo {author} {\bibfnamefont {S.}~\bibnamefont
  {Ospelkaus}},\ }\bibfield  {title} {\bibinfo {title} {Characterizing the
  zeeman slowing force for 40ca19f molecules},\ }\href
  {https://doi.org/10.1088/1367-2630/ac1ed7} {\bibfield  {journal} {\bibinfo
  {journal} {New J. Phys.}\ }\textbf {\bibinfo {volume} {23}},\ \bibinfo
  {pages} {093013} (\bibinfo {year} {2021})}\BibitemShut {NoStop}%
\end{thebibliography}%

\end{document}